\documentclass[sigconf,table,xcdraw]{acmart}

\usepackage{amsmath,amsfonts}
\usepackage{algorithmic}
\usepackage{graphicx}
\usepackage{textcomp}
\usepackage{hyperref}
\usepackage{pifont}
\usepackage{caption}
\usepackage{geometry}
\usepackage{longtable}
\usepackage{array}
\usepackage{booktabs}
\usepackage{enumitem}
\usepackage{multirow}
\usepackage{listings}
\usepackage{todonotes}
\usepackage{tcolorbox}
\usepackage{fancyhdr}
\usepackage{algorithm}
\usepackage{colortbl}

\usepackage{titlesec}

\lstdefinestyle{customc}{
  belowcaptionskip=1\baselineskip,
  breaklines=true,
  frame=L,
  xleftmargin=\parindent,
  language=C,
  showstringspaces=false,
  numbers=left,               % Adds line numbers
  numberstyle=\tiny\color{gray}, % Style for line numbers
  stepnumber=1,
  basicstyle=\scriptsize\ttfamily, % Reduced font size for the code blocks
  keywordstyle=\bfseries\color{green!40!black},
  commentstyle=\itshape\color{purple!40!black},
  numbersep=8pt,
  identifierstyle=\color{blue},
  stringstyle=\color{orange},
}

\titlespacing*{\section}{0pt}{\baselineskip}{0.5\baselineskip} % 

\renewcommand\footnotetextcopyrightpermission[1]{} % Removes the copyright notice

\acmConference[]{}{}{} % Clears the conference information block

\begin{document}

\title{VulnLLMEval: A Framework for Evaluating~Large~Language~Models in Software~Vulnerability~Detection~and~Patching}

\author{Arastoo Zibaeirad}
\affiliation{%
  \institution{University of North Carolina at Charlotte}
  \city{Charlotte}
  \state{NC}
  \country{USA}
}
\email{azibaeir@charlotte.edu}

\author{Marco Vieira}
\affiliation{%
  \institution{University of North Carolina at Charlotte}
  \city{Charlotte}
  \state{NC}
  \country{USA}
}
\email{marco.vieira@charlotte.edu}

\begin{abstract}
Large Language Models (LLMs) have shown promise in tasks like code translation, prompting interest in their potential for automating software vulnerability detection (SVD) and patching (SVP). To further research in this area, establishing a benchmark is essential for evaluating the strengths and limitations of LLMs in these tasks. Despite their capabilities, questions remain regarding whether LLMs can accurately analyze complex vulnerabilities and generate appropriate patches. This paper introduces VulnLLMEval, a framework designed to assess the performance of LLMs in identifying and patching vulnerabilities in C code. Our study includes 307 real-world vulnerabilities extracted from the Linux kernel, creating a well-curated dataset that includes both vulnerable and patched code. This dataset, based on real-world code, provides a diverse and representative testbed for evaluating LLM performance in SVD and SVP tasks, offering a robust foundation for rigorous assessment. Our results reveal that LLMs often struggle with distinguishing between vulnerable and patched code. Furthermore, in SVP tasks, these models tend to oversimplify the code, producing solutions that may not be directly usable without further refinement.
\end{abstract}

% \begin{IEEEkeywords}
% Large Language Models (LLMs), Cybersecurity, Software Vulnerability Detection, Patching
% \end{IEEEkeywords}

\maketitle

\section{Introduction}

% Context, Importance, Introduction, and potential of LLMs in SVD and SVP and motivations
Large Language Models (LLMs), such as OpenAI's Codex \cite{chen2021evaluating}, Meta's Llama \cite{touvron2023llama, dubey2024llama}, CodeBERT \cite{feng2020codebert}, DeepMind's AlphaCode \cite{li2022competition}, Salesforce's CodeGen \cite{nijkamp2022codegen} and CodeT5 \cite{wang2021codet5}, PLBART \cite{ahmad2021unified}, and InCoder \cite{fried2022incoder}, have shown remarkable capabilities in understanding programming languages \cite{nam2024using}. These models have potential for assisting developers by providing code suggestions, test generation \cite{alshahwan2024automated, wang2024software, schafer2023empirical}, code documentation \cite{khan2022automatic}, and automating various coding tasks. In fact, by employing advanced methods to train billions of parameters on data from large datasets \cite{hoffmann2022trainingcomputeoptimallargelanguage, gholami2024improvedgeneralizationboundscommunication, digest, thomas2022dynamic}, LLMs demonstrate significant capabilities in various areas \cite{fan2023large}. These models show great potential for automating Software Vulnerability Detection (SVD) and Software Vulnerability Patching (SVP), making software secure by improving efficiency, accuracy, and coverage in identifying and fixing vulnerabilities.

The number of identified and patched software vulnerabilities has surged in recent years, with over 29,000 Common Vulnerabilities and Exposures (CVEs) addressed in 2023, up from 25,084 in 2022 and 20,153 in 2021 \cite{cvedetails2024}. This trend underscores the need for automated SVD and SVP to keep up with rapidly evolving digital technologies.
There are, however, several limitations and challenges in SVD and SVP from the perspective of both LLMs and traditional vulnerability detection techniques, such as Automatic Program Repair (APR), fuzzing, and Static Analysis Tools (SAT).

From the perspective of LLMs, one major challenge lies in understanding the inherent complexity of vulnerability patterns, which involve intricate interactions between multiple functions or modules within the entire system rather than isolated code snippets. For instance, a vulnerability might only surface when a specific sequence of function calls is made, or when data flows through various parts of the code in a particular order. Detecting these interactions requires a deep understanding of how different parts of the code interact over time.

 A key problem with state-of-the-art LLMs is that they are trained on billions of lines of code without distinguishing between vulnerable and non-vulnerable code, which can lead to ineffective identification and prevention of software vulnerabilities. For instance, while many LLMs are trained on data including the Linux kernel project and know the entire codebase, they lack awareness of previous versions of the Linux kernel that contained numerous vulnerabilities. Furthermore, from a software engineering perspective, it is crucial that these models can detect vulnerabilities in code that they have never seen before, such as new projects or entirely new codebases. To enhance their effectiveness in SVD and SVP, a large dataset that differentiates between vulnerable code versions and their patched counterparts is necessary.

Traditional vulnerability detection techniques also present significant challenges. APR faces issues with patch quality and generalization \cite{huang2023survey, zhang2023survey}. Fuzzing struggles with achieving comprehensive coverage and detecting logic errors \cite{manes2019art, klees2018evaluating}. Similarly, SATs often suffer from a high rate of false positives and difficulty in understanding dynamic behaviors \cite{pereira2021machine}. These limitations highlight the need for development of more sophisticated approaches, including  effectively leveraging LLMs for SVD and SVP, which in turn requires a better understanding of their performance.

% Need for Benchmarking
There are, however, notable gaps in the evaluation of LLMs for SVD and SVP tasks:
\begin{enumerate}
    \item \textbf{Data leakage:}
    The evaluation of state-of-the-art LLMs is compromised by data leakage, where models are tested on datasets they were trained on, leading to inflated performance metrics that do not reflect real-world capabilities \cite{sallou2024breaking, wu2023effective}.
    
    \item \textbf{Comprehensive evaluation with real-world data:}
    Existing evaluations often assess LLMs in narrow contexts using synthesized code snippets, rather than broader real-world scenarios, which limits the realism and scope of the assessments. Incorporating a wider range of real-world code would provide more accurate and robust evaluations, better reflecting LLM performance in practical applications \cite{chakraborty2021deep, wu2023effective}.

    \item \textbf{Manual dataset adjustments:}
    Current methods often rely on manual ground truth patching, limiting full automation of the evaluation process and reducing the number of samples that can be efficiently used \cite{russell2018automated, le2024automatic}.

    \item \textbf{Dynamic and real-time adaptation:}
    adapting LLMs with up-to-date data is essential for effective SVD and SVP. Keeping models current with the latest threats and vulnerabilities enhances their effectiveness and reliability \cite{chang2024survey}.
\end{enumerate}

In this context, the primary objective of a benchmark should be to evaluate the effectiveness and robustness of LLMs in SVD and SVP tasks, aiming to assess how well these models can identify and fix vulnerabilities in real-world codebases \cite{sarker2023machine}. This study aims to fill these gaps by including reproducible real-world vulnerabilities with detailed and realistic assessments, ensuring robust evaluations and practical applicability.

% Purpose and Objective of the Study
We propose VulnLLMEval, a framework to benchmark the performance of pre-trained LLMs in SVD and SVP. Our framework is used to assess ten LLMs and includes a dataset of 307 real-world code vulnerabilities in C code, covering 30 different CWEs, extracted from the Linux kernel. We chose the C language due to its rich dataset availability, the complexity and challenge posed by its low-level constructs, and its critical impact in high-stakes applications. Since publicly available datasets often introduce bias due to data leakage and fail to fully capture real-world vulnerabilities, we developed a novel automated data collection method that gathers and labels vulnerable and patched code using commit hashes. This method eliminates the need for security experts to manually provide ground truth labels, streamlining the evaluation process. By leveraging commit histories from repositories like the Linux kernel, we can systematically extract code changes associated with vulnerabilities and their subsequent patches.

The benchmark is designed for easy extension, enabling the inclusion of additional code samples, evaluation methods, and LLMs to broaden its scope. We employ diverse metrics such as Mean Reciprocal Rank (MRR) and Top-5 accuracy for SVD, and Rouge score, CodeBLEU, and Cyclomatic Complexity for SVP. For binary classification tasks in SVD (e.g., identifying vulnerable code), we also use accuracy, precision, recall, and F1 score. Eight tailored prompts guide LLMs through tasks such as identifying CVEs, CWEs, and code patching.

% Research Questions
\textbf{Research Questions.} This study provides a benchmark designed to explore key research questions as follows:
\newline
\textbf{RQ1:} How effective are pre-trained LLMs in SVD in real-world codebases?
\newline
\textbf{RQ2:} How well do pre-trained LLMs perform in SVP in real-world codebases, and what is the impact of zero-shot versus few-shot learning on their performance?
\newline
\textbf{RQ3:} How do the abstraction levels—ranging from single files with single functions to multiple files with multiple functions—and the length of code blocks affect the accuracy of SVD?

% Key Findings
% \textbf{Key Findings}
% Contributions
\vspace{0.5mm}
\textbf{Contributions.} The main contributions of this study including:
\begin{itemize}
    \item A novel data collection method that makes the process easier for researchers by eliminating the need for security experts to provide ground truth patches for vulnerable code, streamlining the evaluation process and increasing accessibility for future research. 
    \item A comprehensive and fully automatic benchmarking framework for LLMs in SVD and SVP, which is designed to be easily extendable to include additional code samples, programming languages, and LLMs.
    \item An empirical evaluation of ten widely-used, open-source LLMs on real-world vulnerabilities.
    \item Detailed insights into the strengths and limitations of LLMs for SVD and SVP tasks, along with proposed research directions to guide future advancements and improvements in this field.
\end{itemize}

The remainder of this paper is structured as follows: \textbf{Section \ref{section:related_work}} reviews related work in benchmarking for LLMs and data collection methodologies.
    \textbf{Section \ref{section:approach}} details our approach to data collection and the evaluation process for existing LLMs.
    \textbf{Section \ref{section:result}} presents the results, including observations and answers to the research questions.
    \textbf{Section \ref{section:discussion}} addresses the challenges and limitations of our study and explores future research directions and potential improvements to enhance this study.
    \textbf{Section \ref{section:threadstovalidity}} outlines the threats to validity in our research. Finally,
    \textbf{Section \ref{section:conclusion}} concludes the study.

\section{Background and Related Work}
\label{section:related_work}
\textbf{Benchmarking LLMs.} Ullah et al. \cite{ullah2024llms} evaluated eight LLMs on their ability to detect and reason about security vulnerabilities using 228 code scenarios, including 30 real-world examples in C and Python. Despite extensive testing, the study found current LLMs to be inconsistent and unreliable for automated vulnerability detection.

Khare et al. \cite{khare2023understanding} examined state-of-the-art LLMs like GPT-4 and CodeLlama, focusing on their ability to identify security vulnerabilities across Java and C/C++ projects. The study highlighted LLMs' struggle with complex, context-dependent vulnerabilities and the need for specialized fine-tuning strategies.

Sun et al. \cite{sun2024llm4vuln} developed a framework for evaluating LLMs' reasoning capabilities in detecting smart contract vulnerabilities. The study emphasized the importance of context and knowledge integration for improving performance in complex scenarios.

Liu et al. \cite{liu2024vuldetectbench} benchmarked 17 LLMs across various tasks using datasets from projects like Apache Subversion and synthetic sources. The study stressed the significance of context and prompt design in enhancing LLM performance.

Gao et al. \cite{gao2023far} introduced VulBench, a benchmark aggregating data from CTF challenges and other high-quality sources, evaluating LLMs on their ability to detect vulnerabilities and root causes in both binary and multi-class classification tasks.

\textbf{Data Collection.}
The \textit{CVEfixes} dataset \cite{bhandari2021cvefixes} automatically compiles vulnerabilities from the National Vulnerability Database, linking them with corresponding fixes in open-source repositories. It includes both vulnerable and patched code samples, enhanced with metadata like programming languages and code metrics. However, it primarily focuses on modified code snippets rather than providing the full context of the surrounding code.

The \textit{Devign} dataset \cite{zhou2019devign} was manually labeled, featuring functions from major C-language projects such as the Linux Kernel and Wireshark. With over 27,000 functions labeled through extensive manual effort, this dataset provides a detailed representation of code, including abstract syntax trees and control flow graphs.

\textit{CodeXGLUE} \cite{lu2021codexglue} aggregates datasets from various open-source projects to create a comprehensive benchmark for code understanding and generation. It includes datasets like BigCloneBench and Devign, supporting a wide array of programming tasks for machine learning models.

The \textit{REVEAL} dataset \cite{chakraborty2021deep} focuses on collecting vulnerabilities from the Linux Debian Kernel and Chromium projects, emphasizing real-world scenarios by providing labeled vulnerable and clean versions of functions. It offers a realistic dataset for training models on vulnerability detection and patching.
\\
Compared with previous works, our study introduces several unique methodologies and improvements. We utilize a comprehensive real-world dataset of 307 vulnerabilities specifically in the C language, with a focus on the Linux kernel project due to its extensive usage and rich history of vulnerabilities. Although our study primarily centers on the Linux kernel, our benchmark is designed to be easily extendable to include other projects, making it adaptable to a wider range of software environments. Our dataset collection process minimizes manual intervention by automating the extraction of vulnerable and patched code blocks, ensuring a broad and realistic evaluation. Unlike existing works that rely on manual data labeling, our fully automated framework eliminates the need for security experts for ground truth labeling, enhancing consistency and reducing biases.\\
Our evaluation covers 8 prompts designed to assess the effectiveness of LLMs in both SVD and SVP tasks, including zero-shot and few-shot learning scenarios. We also provide a detailed analysis of complex, context-dependent vulnerabilities, offering insights into intricate interactions within the code. Additionally, unlike existing datasets, which often present a narrow view of vulnerabilities, our approach ensures that entire function and non-function elements affected by each commit are extracted and categorized, offering a holistic view of the code changes. A broad range of evaluation metrics, including Mean Reciprocal Rank (MRR), Top-5 accuracy, Rouge score, CodeBLEU, and Cyclomatic Complexity, are integrated to offer a comprehensive assessment. Table \ref{table:related_work} compares our contributions with five recent papers based on several key criteria, highlighting the distinctiveness and robustness of our approach.
\begin{table}[h!]
    \centering 
    \captionsetup{skip=4pt}
    \captionof{table}{Comparison with Existing Work}
    \label{table:related_work}
    \resizebox{\columnwidth}{!}{
    \begin{tabular}{lcccccc}
    \toprule
    \textbf{Key Criteria} & \textbf{Our Paper} & \textbf{\cite{ullah2024llms}} & \textbf{\cite{khare2023understanding}} & \textbf{\cite{sun2024llm4vuln}} & \textbf{\cite{liu2024vuldetectbench}} & \textbf{\cite{gao2023far}} \\ 
    \midrule
    Real-World Dataset & {\color{green}\ding{51}} & {\color{orange}\ding{115}} & {\color{orange}\ding{115}} & {\color{red}\ding{55}} & {\color{green}\ding{51}} & {\color{orange}\ding{115}} \\
    Novel Data Collection & {\color{green}\ding{51}} & {\color{red}\ding{55}} & {\color{red}\ding{55}} & {\color{red}\ding{55}} & {\color{red}\ding{55}} & {\color{orange}\ding{115}} \\ 
    Automated Evaluation & {\color{green}\ding{51}} & {\color{green}\ding{51}} & {\color{orange}\ding{115}} & {\color{orange}\ding{115}} & {\color{green}\ding{51}} & {\color{red}\ding{55}} \\ 
    No Security Experts for Labeling & {\color{green}\ding{51}} & {\color{red}\ding{55}} & {\color{red}\ding{55}} & {\color{red}\ding{55}} & {\color{green}\ding{51}} & {\color{green}\ding{51}} \\ 
    Multiple Prompt Evaluation & {\color{green}\ding{51}} & {\color{green}\ding{51}} & {\color{green}\ding{51}} & {\color{green}\ding{51}} & {\color{orange}\ding{115}} & {\color{orange}\ding{115}} \\ 
    Holistic Metrics Integration & {\color{green}\ding{51}} & {\color{green}\ding{51}} & {\color{orange}\ding{115}} & {\color{orange}\ding{115}} & {\color{orange}\ding{115}} & {\color{orange}\ding{115}} \\ 
    \bottomrule
    \end{tabular}
    }
    \begin{flushleft} % Left-align this section
    \centering
    {\color{green}\ding{51}}Fully addressed {\color{orange}\ding{115}}Partially addressed  {\color{red}\ding{55}}Not addressed
    \vspace{0.3cm}
    \end{flushleft}
\end{table}
% \vspace{2cm}
\section{Benchmarking Approach}
\label{section:approach}
In this section, we present our framework and methodology to evaluate the performance of LLMs in SVD and SVP. We start by introducing VulnLLMEval in Section \ref{subsection:framework}, outlining the steps and tools we used in our evaluation. Section \ref{subsection:dataset} then describes the dataset and how we extracted real-world vulnerabilities from the Linux kernel, with potential extensions to other projects in the future. In Section \ref{subsection:metrics}, we cover the different metrics used to measure the effectiveness of LLMs in both SVD and SVP tasks, ensuring a thorough and balanced evaluation. Finally, Section \ref{subsection:prompts} explains the prompt templates created to standardize the input queries for the LLMs.

\subsection{VulnLLMEval}
\label{subsection:framework}
\begin{figure*}[htbp]
\centerline{\includegraphics[width=1\textwidth]{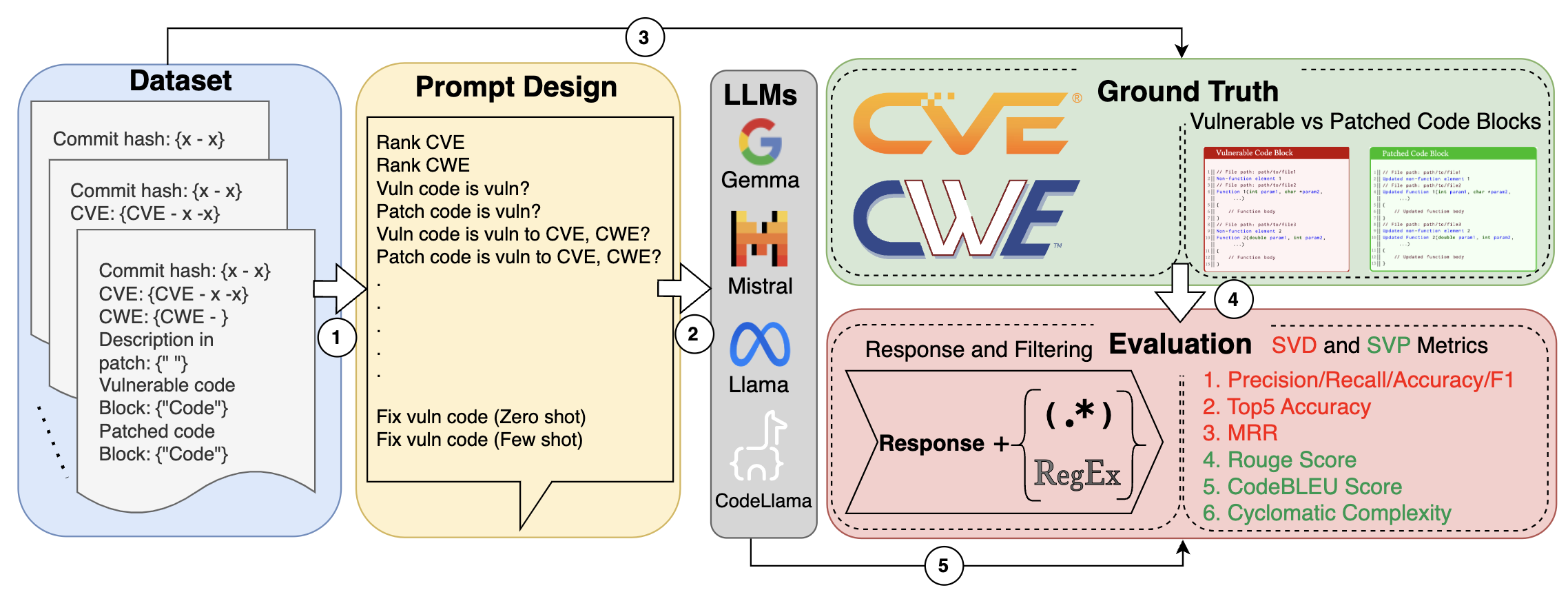}}
\caption{VulnLLMEval Architecture}
\label{figure:framework}
\end{figure*}

VulnLLMEval is a framework designed to evaluate the effectiveness and robustness of LLMs in tasks related to SVD and SVP, as shown in Figure \ref{figure:framework}. The process starts with the extraction of initial metadata, such as commit hashes linked to vulnerabilities with CVE and CWE identifiers, forming the foundation for building our dataset of vulnerable and patched code blocks, as discussed in Section \ref{subsection:dataset}. The systems under benchmarking (SUB) are ten LLMs pre-trained on various programming languages and code structures (detailed in Table \ref{table:llms}), and VulnLLMEval evaluates their performance across 8 tasks to benchmark how effectively these models can identify and patch vulnerabilities in real-world codebases. The outputs generated by the LLMs are automatically filtered using regular expressions, ensuring consistency and reducing biases or errors. Finally, the evaluation involves calculating precision, recall, accuracy, F1 scores, and other relevant metrics for detection, along with ROUGE, CodeBLEU, and cyclomatic complexity for patching, as detailed in Section \ref{subsection:metrics}.

\textbf{Scope of the Benchmark.} VulnLLMEval targets C/C++ code, given its widespread use in system-level programming, particularly in projects like the Linux kernel, one of the most widely used and essential systems globally. Due to its prevalence and the critical nature of its vulnerabilities, the Linux kernel serves as a highly relevant target for our benchmark. While this benchmark focuses on the Linux kernel, it can be extended to other programming languages. It considers over 30 CWEs, with an emphasis on prevalent vulnerabilities such as buffer overflows (CWE-119), improper input validation (CWE-20), information exposure (CWE-200), use-after-free (CWE-416), and null pointer dereference (CWE-476). The evaluation assesses code at both the file and function levels, incorporating non-functional elements like global variables and configuration settings linked to functional components. Including both functional and non-functional elements is essential as it captures the interconnections between files, functions, and non-functional components, providing a comprehensive view of the codebase. This approach is crucial for understanding the complexity of vulnerable code since patches often involve changes to variables or settings that LLMs must accurately interpret for effective vulnerability detection and patching.

\textbf{Evaluation Scenarios and Metrics.} VulnLLMEval can be used to assess LLMs through two primary scenarios: detection and patching. In the detection scenario, the models are evaluated on their ability to identify vulnerabilities within given code blocks using metrics such as precision, recall, accuracy, F1 score, top-K accuracy, and mean reciprocal rank (MRR). In the patching scenario, the effectiveness of the LLMs in generating patches for identified vulnerabilities is measured using ROUGE score, CodeBLEU score, and the cyclomatic complexity of the generated patches. Further details on these metrics are provided in Section \ref{subsection:metrics}.

\textbf{Workload.} The workload comprises a diverse set of real-world vulnerable and patched code blocks from the Linux kernel project, selected based on diversity, representativeness, and complexity. This ensures a comprehensive evaluation of the performance of the models across various levels of difficulty. In practice, the workload is tailored for different scenarios:
\begin{itemize}
    \item \textbf{Detection:} This workload is designed to test the accuracy and efficiency of LLMs in recognizing and flagging vulnerabilities within Linux kernel source code. It focuses on the ability of LLMs to identify vulnerabilities from a set of given code blocks.
    \item \textbf{Patching:} This workload focuses on generating patches for vulnerable code blocks to evaluate the effectiveness of LLMs in providing accurate fixes. The patches are analyzed for complexity, similarity, and the extent of line changes to understand the LLM's capability in handling various levels of code modifications.
\end{itemize}

To ensure the reliability of our benchmark, we focus on several key properties. We prioritize \textit{representativeness} by including a diverse array of real-world vulnerabilities and code structures, making the dataset as reflective as possible of practical scenarios. Although our dataset collection process is tailored for the C language—using regular expressions specifically designed for C—the underlying approach is adaptable and can be extended to other programming languages. We also demonstrate the \textit{portability} of the benchmark by successfully applying it across various LLMs, showcasing its flexibility and broad applicability. \textit{Fairness} is a cornerstone of our approach achieved by using standardized, widely accepted evaluation metrics, and consistently applying the same dataset and procedures across all models for unbiased and meaningful comparisons.

\subsection{Dataset}
\label{subsection:dataset}
Manually labeling datasets is labor-intensive, time-consuming, and difficult to scale, often leading to incomplete or inconsistent data \cite{le2024automatic}, so we developed a custom dataset to address these challenges efficiently. Our approach focuses on creating datasets that distinguish between vulnerable and non-vulnerable code in real-world scenarios, rather than just using isolated code snippets of vulnerable code. By leveraging publicly available CVE records and metadata such as commit hashes, CVEs, and CWEs, we can systematically extract vulnerable code blocks and their corresponding patched versions at various levels of abstraction, including files, functions, and non-functional elements. This ensures a comprehensive understanding of vulnerabilities and their contexts, providing a more realistic and effective dataset for training and evaluating LLMs. The detailed steps of this automated data collection process are outlined in Algorithm \ref{alg:vulnerable_code_collection}.

\begin{algorithm}
\caption{Vulnerable/Patched Code Block and Metadata Collection}
\label{alg:vulnerable_code_collection}
\begin{algorithmic}[1]
\REQUIRE Repository URL $R$, List of Commits $C$ with Commit Hashes
\ENSURE Database $D$ with Vulnerable and Patched Code Blocks

\FOR{each commit $c \in C$}
    \STATE Fetch the patch from $R$ using $c$.
    \STATE Parse the patch to extract:
    \begin{itemize}
        \item Files $F$
        \item Functions $F_{func}$
        \item Added Lines $L_{add}$
        \item Deleted Lines $L_{del}$
    \end{itemize}
\ENDFOR

\FOR{each change $l \in (L_{add} \cup L_{del})$}
    \IF{$l$ is inside a function}
        \STATE Retrieve the function before fix $F_{func}^-$ at parent commit.
        \STATE Retrieve the function after fix $F_{func}^+$ at current commit.
        \STATE Append $F$ and $F_{func}$ to the Vulnerable/Patched Code Blocks in $D$.
    \ELSE
        \STATE Append $F$ and $l$ to the Non-functional Code Blocks in $D$.
    \ENDIF
\ENDFOR

\IF{a function pattern is identified within $l$}
    \STATE Retrieve $F_{func}^-$ and $F_{func}^+$ as in Step 2.1.
    \STATE Update $D$ with these functions.
\ENDIF

\STATE Store the following in database $D$:
\begin{itemize}
    \item Vulnerable Code Blocks
    \item Patched Code Blocks
    \item Associated Metadata
\end{itemize}

\end{algorithmic}
\end{algorithm}

\textbf{Dataset Acquisition and Preparation.} 
Our study centers on the Linux kernel project, which offers a wealth of vulnerability data crucial for refining our data collection process. We systematically gathered essential metadata, including commit hashes, CVEs, and CWEs, utilizing resources introduced by Pereira et al. \cite{pereira2022software}. Additionally, metadata was sourced from platforms such as MITRE \cite{mitre_cve} and the National Vulnerability Database (NVD) \cite{nist_nvd}. This approach enabled us to trace the history and progression of vulnerabilities—from their initial discovery to their resolution—providing a comprehensive understanding of how vulnerabilities evolve and are patched over time. The output of this step is the collection of commit hashes linked to specific CVEs and their corresponding CWEs, ensuring a detailed and accurate dataset for further analysis. This comprehensive gathering process allows for a thorough analysis of the Linux kernel’s codebase, capturing the complete lifecycle of each vulnerability from introduction to remediation.

\textbf{Extraction of Vulnerable and Patched Codes.} Once the initial dataset was acquired, we focused on extracting detailed code changes for each commit. Leveraging commit hashes along with their corresponding file paths allowed us to retrieve the vulnerable code from the parent commit and the patched code from the commit itself with precision. Each commit was analyzed for comprehensive patch information, including file changes, file paths, line modifications, and specific alterations.

\textbf{Function and Non-Function Elements Extraction.} Following the retrieval of code blocks, we focused on identifying and extracting both the functions and non-functional elements that were modified by each commit. To do this, we parsed the patch files to precisely identify the function names and the specific changes made using advanced regular expressions. However, a challenge we encountered was that patch files sometimes referenced function names that were not directly modified, often to provide additional context about adjacent changes. To address this, we ensured that only functions with added or deleted lines directly within them were extracted as vulnerable or patched functions. If no direct changes were made within a function, we relied on regular expression patterns to identify function names correctly within the code changes. Lines of code that were not part of specific functions-such as libraries, global variables, macros, and type definitions-were grouped separately. This approach offers a complete view of each change, ensuring a detailed and well-organized representation of the codebase.

\textbf{Commit Descriptions.} To enhance LLMs' few-shot learning capabilities in SVP tasks, we extracted detailed commit descriptions that provide crucial context for each vulnerability. These descriptions include the rationale behind code changes and specifics of the modifications made to address the vulnerabilities. The rationale offers insights into the problem being solved, while the modification details explain how the vulnerability was patched. This contextual information helps LLMs better understand the purpose of the changes and supports them in generating more accurate fixes. However, to avoid introducing bias in SVD tasks, we did not use commit descriptions, as they sometimes explicitly mention the vulnerability details. Comments within the code were retained for both SVD and SVP tasks, and these were used in both zero-shot and few-shot learning scenarios. This approach ensures that the LLMs receive relevant contextual information without compromising the integrity of the evaluation.

\textbf{Comprehensive Code Blocks Assembly.} To provide a holistic view of each vulnerability, we compiled all modified functions and non-functional elements into distinct blocks, creating a structured representation of both vulnerable and patched code states. This approach addresses the limitations of existing datasets, which often provide an incomplete view by focusing on isolated snippets or small code fragments without considering the broader context of changes. By ensuring that LLMs receive a complete picture of the code changes—including functional and non-functional modifications—this method allows for a more detailed and realistic evaluation of their capabilities in detecting and patching vulnerabilities, enhancing the overall reliability of our evaluation framework.

\textbf{Updating and Maintaining the Dataset.} Recognizing the dynamic nature of software development and the continuous emergence of new vulnerabilities, we designed our dataset collection process to support ongoing updates. Our methodology facilitates the periodic collection of new data, ensuring that the dataset remains current and accurately reflects the evolving security landscape. The process requires access to commit hashes for code retrieval and CVE and CWE identifiers for ground truth labeling in subsequent evaluation steps. By utilizing commit hashes, we can continuously extract both the vulnerable and corresponding patched code blocks, maintaining a comprehensive and up-to-date dataset for future research and evaluation. The code is available in our GitHub repository\footnote{The repository will be made publicly available after acceptance by the conference.}.

In our study, we extract and organize vulnerable and patched code blocks to assess how effectively LLMs can identify and patch software vulnerabilities. We structure each code block by its file paths, clearly distinguishing between non-functional elements and core function bodies. As exemplified by Listing \ref{code-blocks}, the original vulnerable code block is juxtaposed with its corresponding patched version, which provides a straightforward, side-by-side comparison. Within each vulnerable code block, file paths are organized with their associated functions and non-functional elements. The full functions, whether vulnerable or patched, are extracted and placed in their respective blocks. For non-functional elements, lines deleted from the code are included in the vulnerable block, while added lines are placed in the corresponding patched block. In cases where an entire function or non-function element was either added or removed, the relevant content is assigned to either the vulnerable or patched block accordingly. This method ensures that all changes, whether in function bodies or non-functional code, are captured and evaluated accurately.  

Our dataset, is both diverse and representative, encompassing 30 unique CWEs. It includes 285 unique CVEs, such as CVE-2005-4881, CVE-2009-2695, and CVE-2019-11487, each appearing multiple times but in different commit hashes and code blocks. Vulnerable code blocks vary widely, averaging 202 lines, with a range from 3 to 3,581 lines, while patched blocks average 209 lines, spanning from 3 to 3,613 lines. This variation in code block size and the wide array of CWEs and CVEs underscores the dataset's comprehensive nature, accurately reflecting real-world software vulnerability challenges. Importantly, the data consists of vulnerabilities across multiple versions of the Linux kernel, rather than focusing on a single version, ensuring comprehensive coverage of vulnerabilities and patches over time. While the dataset collection process is currently based on advanced regular expressions tailored for C/C++, this approach can be extended to other programming languages by adapting language-specific regular expressions, enabling the framework to handle vulnerabilities across a broader range of software contexts.
\vspace{0.2cm}
% \lipsum[1]

\setlength{\fboxsep}{0pt}
\label{code-blocks}

\noindent
\begin{minipage}[t]{0.47\columnwidth}
  \begin{tcolorbox}[colback=red!5!white, colframe=red!75!black, title=Vulnerable Code Block, equal height group=A, height=5.5cm]
    \lstset{style=customc}
    \begin{lstlisting}
// File path: path/to/file1
Non-function element 1
// File path: path/to/file2
Function 1(int param1, char *param2, ...)
{
    // Function body
}
// File path: path/to/file3
Non-function element 2
Function 2(double param1, int param2, ...)
{
    // Function body
}
    \end{lstlisting}
  \end{tcolorbox}
\end{minipage}
\hfill
\begin{minipage}[t]{0.52\columnwidth}
  \begin{tcolorbox}[colback=green!5!white, colframe=green!75!black, title=Patched Code Block, equal height group=A, height=5.5cm]
    \lstset{style=customc}
    \begin{lstlisting}
// File path: path/to/file1
Updated non-function element 1
// File path: path/to/file2
Updated Function 1(int param1, char *param2, ...)
{
    // Updated function body
}
// File path: path/to/file3
Updated non-function element 2
Updated Function 2(double param1, int param2, ...)
{
    // Updated function body
}
    \end{lstlisting}
  \end{tcolorbox}
\end{minipage}

% \lipsum[2]

\subsection{Metrics}
\label{subsection:metrics}
To evaluate our models, we employ various metrics that characterize the accuracy and relevance of predictions. These metrics are tailored for both SVD and SVP.
\subsubsection{SVD Metrics}
\begin{enumerate}
    \item \textbf{F1 Score, Recall, Accuracy, Precision} are standard metrics for evaluating the performance of classification models. Accuracy measures the ratio of correctly predicted instances to the total instances. Precision is the ratio of correctly predicted positive observations to the total predicted positives. Recall (Sensitivity) is the ratio of correctly predicted positive observations to all actual positives. The F1 Score, the harmonic mean of precision and recall, provides a single metric that balances both precision and recall. These metrics are fundamental for assessing the effectiveness of binary and multi-class classification tasks.

    \begin{align} 
    \scriptsize
        &\text{Accuracy} = \frac{\text{TP} + \text{TN}}{\text{TP} + \text{TN} + \text{FP} + \text{FN}}\\
        &\text{Precision} = \frac{\text{TP}}{\text{TP} + \text{FP}}\\
        &\text{Recall} = \frac{\text{TP}}{\text{TP} + \text{FN}}\\
        &\text{F1 Score} = 2 \times \frac{\text{Precision} \times \text{Recall}}{\text{Precision} + \text{Recall}}
    \end{align}

    \item \textbf{Top-k accuracy} is a metric used to evaluate classification models where the model's prediction is considered correct if the true label is within the top \(k\) predicted labels. This metric is particularly useful in scenarios where ranking is crucial, such as recommendation systems, where it is important to measure how often the correct answer is among the top predictions. In this study, we specifically use Top-5 accuracy, as it provides a balanced assessment of model performance by focusing on the most likely correct answers while keeping the evaluation manageable \cite{russakovsky2015imagenet}. The equation is:

    \begin{equation}
     \frac{\text{Number of correct predictions in top k}}{\text{Total number of predictions}}
    \end{equation}

    \item \textbf{Mean Reciprocal Rank (MRR)} \cite{craswell2009mean} is a statistical measure commonly used in information retrieval and question-answering systems to assess how effectively a system ranks the correct answers. In the context of LLMs, MRR evaluates the quality of responses in classification tasks by measuring how high the correct answer is positioned in a ranked list of possible answers based on their probabilities. MRR provides an average score that reflects the ability to rank relevant items at the top.

\begin{equation}
    \text{MRR} = \frac{1}{N} \sum_{i=1}^{N} \frac{1}{\text{rank}_i}
\end{equation}

Here, \( N \) is the number of queries, and \(\text{rank}_i\) is the rank position of the first relevant document or correct answer for the \( i \)-th query.

\end{enumerate}

\subsubsection{SVP Metrics}
\begin{enumerate}

\item \textbf{ROUGE Score} \cite{lin2004rouge} is used to evaluate the similarity between the generated patches and reference patches by comparing their n-gram overlap. ROUGE-L, which focuses on the longest common subsequence, is particularly relevant for measuring how well the generated patch aligns with the reference in terms of sequence structure.

In practice, after generating a patch, we calculate the ROUGE-L score by comparing it to the reference patch. A higher ROUGE-L score indicates that the model has produced a patch that closely matches the reference in terms of sequence similarity, which can be an indicator of the model capturing relevant patterns.

\begin{equation}
\text{ROUGE-L} = \frac{\sum_{\text{LCS}(X,Y)}}{\sum_{\text{Count}(X)}} 
\end{equation}

where \(\text{LCS}(X,Y)\) represents the length of the longest common subsequence between the generated text \(X\) and reference text \(Y\), and \(\text{Count}(X)\) is the total length of the reference text.

\item \textbf{CodeBLEU} \cite{ren2020codebleu} is an advanced metric designed specifically for evaluating code generation. Traditional metrics like BLEU \cite{papineni2002bleu} are inadequate for code as they primarily focus on n-gram matching, overlooking essential aspects such as syntactic structure and semantic correctness. While metrics like ROUGE can capture surface-level similarities, they fall short in evaluating code since they only measure textual overlap and do not account for the underlying structure or logic. CodeBLEU goes beyond these limitations by assessing not only surface similarity but also the syntactic and semantic accuracy of the generated code, making it a more comprehensive tool for code evaluation.

CodeBLEU combines several dimensions of code quality, including:
\begin{itemize}
    \item \textbf{Surface similarity} through n-gram matching.
    \item \textbf{Syntactic accuracy} using abstract syntax tree (AST) matching.
    \item \textbf{Semantic correctness} via data-flow analysis.
\end{itemize}

The overall CodeBLEU score is calculated as a weighted sum of these components:
\begin{equation}
\text{CodeBLEU} = \alpha \cdot \text{BLEU} + \gamma \cdot \text{Match}_{\text{AST}} + \delta \cdot \text{Match}_{\text{DF}}
\end{equation}
where \(\text{BLEU}\) refers to the standard BLEU score, \(\text{BLEU}_{\text{weight}}\) applies weights to different n-grams, emphasizing the importance of certain tokens, particularly keywords, \(\text{Match}_{\text{AST}}\) evaluates syntactic correctness by comparing sub-trees in the abstract syntax trees (ASTs) of the candidate and reference code, and \(\text{Match}_{\text{DF}}\) assesses semantic accuracy by comparing data-flow graphs, which represent variable dependencies.
This approach ensures that the generated code not only matches the reference text in form but also aligns with it in structure and meaning, making CodeBLEU a more reliable metric for evaluating code generation.

    \item \textbf{Cyclomatic Complexity} \cite{mccabe1976complexity} is a software metric used to measure the complexity of a program by quantifying the number of linearly independent paths through the source code. This metric helps identify complex sections of code that may require more rigorous testing and maintenance, ultimately aiding in managing code quality and maintainability. By comparing the complexity of the original vulnerable code to the patched version, we assess whether the LLMs are generating simplified but functional code or producing overly complex fixes that may introduce maintenance challenges. A decrease in complexity often indicates more maintainable code, while excessive reductions may suggest oversimplifications that overlook important functionality. In our study, the cyclomatic complexity ranges from 0 to 5 (low complexity), 5 to 10 (moderate complexity), 10 to 15 (high complexity), and 15 to 20 (very high complexity), offering a framework for evaluating whether the patches generated by LLMs simplify the code appropriately without compromising functionality.
    
    It is calculated using the formula:
    
    \begin{equation}
        M = E - N + 2P
    \end{equation}

where \(M\) is the cyclomatic complexity, \(E\) is the number of edges representing control flow transitions between nodes in the control flow graph of the program, \(N\) refers to the number of nodes which indicate decision points or statements, and \(P\) is the number of connected components that represent independent sections of code, such as separate functions or methods.

\end{enumerate}

\subsection{Prompt Templates}
\label{subsection:prompts}
In this section, we outline the prompt templates used for zero-shot and few-shot scenarios in vulnerability detection and patching. However, we excluded few-shot prompts for vulnerability detection to prevent bias. Table \ref{table:prompt_templates} summarizes the prompt types, inputs, and descriptions.

\begin{table}[h!]
    \centering
    \captionsetup{skip=4pt}
    \caption{Prompt Templates}
    \resizebox{\columnwidth}{!}{
        \begin{tabular}{p{0.8cm}p{3.5cm}p{2cm}p{6.8cm}} 
        \toprule
        \textbf{ID} & \textbf{Type} & \textbf{Inputs} & \textbf{Description} \\ \midrule
        SVD1 & Rank CVEs (Z) & V & Rank the top 5 probable CVEs for the vulnerable code. \\ 
        SVD2 & Rank CWEs (Z) & V & Same as SVD1, but for CWEs. \\ 
        SVD3 & Is Vulnerable (Z) & V & Identify if the code block is vulnerable. \\ 
        SVD4 & Is Patched Vulnerable (Z) & P & Same as SVD3, but for patched code block. \\ 
        SVD5 & CVE/CWE-Vuln Check (Z) & V, CVE, CWE & Verify if the code is vulnerable to its CVE and CWE. \\ 
        SVD6 & CVE/CWE-Patch Check (Z) & P, CVE, CWE & Same as SVD5, but for patched code block. \\ 
        SVP1 & Suggest a Fix (Z) & V & Provide a fix for the vulnerable code block. \\ 
        SVP2 & Suggest a Fix (F) & V, D & Same as SVP1, but with extra context.\\ \bottomrule
        \end{tabular}
    }
    \begin{flushleft}
    \scriptsize
    Zero-shot: Z; Few-shot: F; Vulnerable code block: V; Patched code block: P; Description: D
    \end{flushleft}
    \label{table:prompt_templates}
\end{table}

Our approach systematically evaluates the LLMs' ability to identify vulnerabilities in through two primary methods: ranking and Yes/No questioning. For CVE and CWE ranking, we use prompts SVD1 and SVD2 in Table \ref{table:prompt_templates}, where the LLMs are provided with vulnerable code and asked to identify the most likely CVEs and CWEs associated with it. The models rank the top five possibilities, and we evaluate their performance using MRR and Top-5 accuracy. MRR measures how highly the correct CVE or CWE is ranked, while Top-5 accuracy checks if the correct vulnerability is in the top five predictions. These metrics quantify the model's ability to prioritize vulnerabilities.

Following the ranking, prompts SVD3 to SVD6 focus on Yes/No questions to assess the models' ability to identify vulnerabilities in a code block. We evaluate performance using accuracy, precision, recall, and F1 Score. In SVD3 and SVD4, the models evaluate whether a code block, whether vulnerable or patched, contains vulnerabilities, while in SVD5 and SVD6, they determine if the block is vulnerable to a specific CVE or CWE provided as the ground truth.

For the patching scenario, we assess the LLMs' ability to generate accurate fixes for vulnerabilities in code blocks using both zero-shot (SVP1) and few-shot (SVP2) settings, as outlined in Table \ref{table:prompt_templates}. In the zero-shot setting, the model generates a patch using only its existing knowledge, testing its independent problem-solving ability. In the few-shot setting, it receives a detailed commit description for additional context. We evaluate the patches using ROUGE, CodeBLEU, and Cyclomatic Complexity to assess their structural and semantic accuracy.

\section{Benchmarking Results}
\label{section:result}

In this study, we leverage 10 pre-trained LLMs to assess their capabilities in SVD and SVP tasks. The models include CodeLlama \cite{roziere2023code}, Gemma2 \cite{team2024gemma}, Llama3, Llama3.1 \cite{dubey2024llama}, and Mistral \cite{jiang2023mistral}, each available in different parameter scales. We primarily used instruct models, as they are fine-tuned to better follow instructions and generate more relevant outputs for complex tasks. While ChatGPT is not included in this study, our primary goal is to demonstrate and validate the benchmark. Our approach is designed to be flexible and can easily be adapted to evaluate other LLMs, including different versions of GPT. The specifications of the models we used are detailed in Table \ref{table:llms}. To maintain consistency across evaluations, we set the temperature parameter to 0.0 for all models.

\begin{table}[ht]
    \centering
    \captionsetup{skip=4pt}
    \caption{LLM Parameters and Specifications}
    \label{table:llms}
    \scriptsize
    \resizebox{\columnwidth}{!}{%
        \begin{tabular}{llcc}
        \toprule
        \textbf{Model Class} & \textbf{Model Version} & \textbf{Context Window Size} \\ 
        \midrule
        \multirow{2}{*}{\textbf{CodeLlama}} 
            & 7B-instruct & 16,384 tokens \\ 
            & 34B-instruct & 16,384 tokens \\
        \midrule
        \multirow{2}{*}{\textbf{Llama3}} 
            & 70B-instruct & 8,192 tokens \\
            & 8B-instruct & 8,192 tokens \\ 
        \midrule
        \multirow{2}{*}{\textbf{Llama3.1}} 
            & 70B & 131,072 tokens \\
            & 8B & 131,072 tokens \\ 
        \midrule
        \multirow{2}{*}{\textbf{Gemma2}} 
            & 27B & 8,192 tokens \\ 
            & 9B & 8,192 tokens \\ 
        \midrule
        \multirow{2}{*}{\textbf{Mistral}} 
            & 7B-instruct  & 32,768 tokens \\ 
            & Mixtral-8*7B & 32,768 tokens \\ 
        \bottomrule
        \end{tabular}
    }
\end{table}
\subsection{RQ1: Effectiveness of LLMs in SVD}
In this section, we examine three critical aspects of evaluating the effectiveness of LLMs in SVD tasks using our real-world dataset. First, we assess the models' ability to accurately rank CVEs and CWEs by comparing their outputs against ground truth data. Next, we analyze how well the models differentiate between vulnerable and patched code through four targeted prompts. Finally, we evaluate the effectiveness in handling the most frequently occurring CWEs, providing detailed insights into their applicability and reliability in real-world SVD tasks. 

\subsubsection{Ranking}
In SVD1 and SVD2, the task is to assess the models' ability to rank CVEs and CWEs accurately. To evaluate this, we calculate Top-5 accuracy and MRR, which measure the precision and relevance of the rankings. MRR provides a more comprehensive evaluation by considering the exact rank of the correct answer, reflecting the overall quality of the models' rankings. To ensure fair benchmarking, we included the year of the CVE in the rankings, preventing confusion from similarly characterized CVEs across different years. In instances where LLMs provided descriptions instead of explicit CVE IDs or varied in the number of ranked responses, we dynamically adjusted the assessment, flagging such responses to avoid skewing the results. Logic was incorporated into the script to ensure that MRR was calculated on valid, comparable rankings, even when response lengths varied.

\textbf{Observations.} Our analysis revealed significant variability in the LLMs' ability to accurately rank CVEs and CWEs. Notably, the performance in ranking CWEs was generally better than for CVEs, potentially due to the broader and more structured nature of CWEs, which may align better with the LLMs' training data. Among the models, Llama3-70b demonstrated the highest performance across both CVE and CWE rankings, consistently placing the correct vulnerabilities higher in the list and achieving the best MRR scores. Conversely, Codellama often ranked the correct vulnerabilities lower on the list, indicating lower accuracy. Additionally, while the models were expected to provide the top 5 most probable CVEs and CWEs, the length of their responses varied—ranging from one to five entries. Despite our inclusion of the year for CVE rankings, some LLMs returned results with CVEs from incorrect years, showing difficulties in maintaining context. Furthermore, certain models provided only descriptions of the vulnerable code without listing CVE IDs, or indicated that they could not generate a list of potential CVEs or lacked sufficient information to identify CWEs in the code. These inconsistencies highlight the challenges in obtaining reliable and consistent outputs from the models. The results of this analysis are summarized in Table \ref{tab:rank-cve-cwe}. 

\begin{figure*}[htbp]
\centerline{\includegraphics[width=1\textwidth]{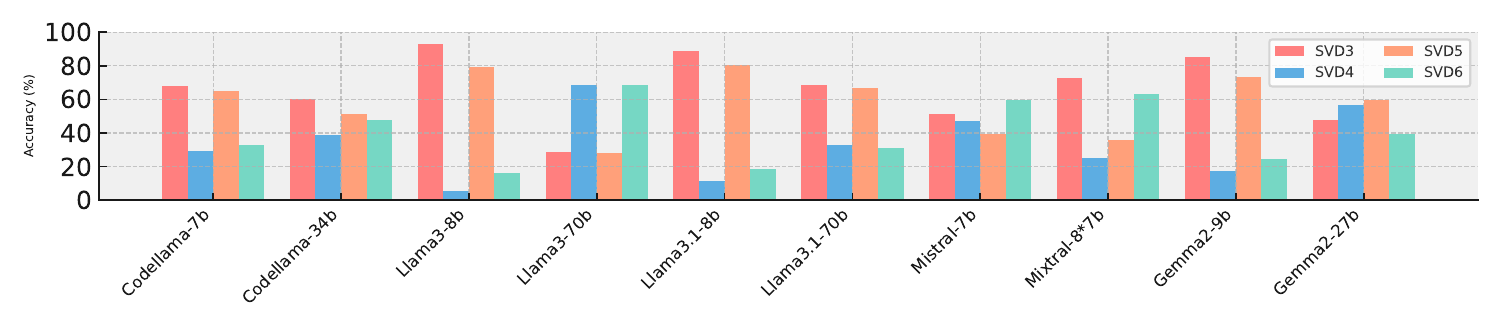}}
\caption{Accuracy of LLMs in Identifying Vulnerable and Patched Code Blocks Across SVD3, SVD4, SVD5, and SVD6 Prompts.}
\label{figure:is-vuln?}
\end{figure*}

\begin{table*}[ht]
    \centering
    \captionsetup{skip=4pt}
    \caption{CVE and CWE Rankings: Top-5 Accuracy and MRR}
    \label{tab:rank-cve-cwe}
    \setlength{\tabcolsep}{4pt} % Adjust space between columns
    \resizebox{\textwidth}{!}{%
    \begin{tabular}{l|lcccccccccc} % Added vertical bars for the data rows only
        \toprule
        \multicolumn{2}{l}{\textbf{Metric}} & \textbf{Codellama-7b} & \textbf{Codellama-34b} & \textbf{Llama3-8b} & \textbf{Llama3-70b} & \textbf{Llama3.1-8b} & \textbf{Llama3.1-70b} & \textbf{Mistral-7b} & \textbf{Mixtral-8*7b} & \textbf{Gemma2-9b} & \textbf{Gemma2-27b} \\
        \midrule
        \multirow{2}{*}{\textbf{Top-5 Accuracy}} & \textbf{CVE} & 0.00 & 0.00 & 0.65 & 5.21 & 0.65 & 8.14 & 0.00 & 0.00 & 0.00 & 0.33 \\
        & \textbf{CWE} & 6.51 & 6.33 & 21.17 & 22.48 & 15.96 & 15.31 & 11.07 & 13.68 & 9.77 & 17.92 \\
        \midrule
        \multirow{2}{*}{\textbf{MRR}} & \textbf{CVE} & 0.00 & 0.00 & 0.23 & 4.29 & 0.23 & 5.79 & 0.00 & 0.00 & 0.00 & 0.16 \\
        & \textbf{CWE} & 3.82 & 3.52 & 9.65 & 12.44 & 9.03 & 7.89 & 4.89 & 6.74 & 7.20 & 9.15 \\
        \bottomrule
    \end{tabular}%
    }
\end{table*}

\subsubsection{Is Vulnerable?}
We assess the effectiveness of LLMs in determining whether a code block is vulnerable. This evaluation focuses on key metrics—precision, recall, accuracy, and F1 score—applied to both vulnerable and patched code blocks across various LLMs. The analysis offers insights into how well these models can distinguish between vulnerable code, patched code (which has been modified to address vulnerabilities). This distinction is crucial for ensuring accurate and reliable security assessments. Specifically, prompts SVD3 and SVD5 focus on code blocks where the ground truth indicates vulnerability (labeled as 1), while SVD4 and SVD6 address code blocks that are not vulnerable (labeled as 0). SVD3 and SVD4 evaluate whether a given code block, either vulnerable or patched, remains vulnerable, while SVD5 and SVD6 challenge the LLMs to determine if the code block is vulnerable based on its corresponding CVE and CWE identifiers. To ensure consistent and accurate evaluations, we implemented a comprehensive set of regular expressions to filter and analyze the LLMs' responses effectively.

\textbf{Observations.}
Table \ref{tab:is-vulnerable} summarizes the results for four evaluation prompts. SVD3 and SVD4 are paired to assess how well the models can identify vulnerabilities, with SVD3 focusing on vulnerable code blocks and SVD4 on patched code blocks. Similarly, SVD5 and SVD6 evaluate the models' ability to determine if a code block is vulnerable by considering whether it is susceptible to its corresponding CVE and CWE identifiers. The combined analysis of SVD3 and SVD4, as well as SVD5 and SVD6, reveals several key trends. Notably, providing the ground truth CVE and CWE in SVD5 and SVD6 results in an average improvement of 9\% in accuracy, indicating that LLMs perform better when given specific contextual information.

However, the models generally tend to classify code as vulnerable regardless of whether the block is the original vulnerable code or its patched version. This tendency explains the higher accuracy observed in SVD3 and SVD5 compared to SVD4 and SVD6, as shown in Figure \ref{figure:is-vuln?}. Interestingly, in the case of the Llama3-70b model, the accuracy for SVD4 and SVD6 was significantly higher than for SVD3 and SVD5, suggesting that this model may handle context differently or better understand the transition between vulnerable and patched states. This anomaly indicates potential areas where certain models might better grasp the nuances between vulnerable and non-vulnerable code when given the right context.

The aggregated accuracy of the LLMs under assessment across each prompt highlights the varying effectiveness of the models: SVD3 achieved an accuracy of 66.51\%, SVD4 had 33.19\%, SVD5 reached 57.92\%, and SVD6 showed 40.13\%. Given these results, we prioritize SVD5 and SVD6 over SVD3 and SVD4 because they assess the model's ability to identify vulnerabilities tied to specific CVEs and CWEs, a more practical and targeted task for real-world scenarios. While SVD3 and SVD4 evaluate general vulnerability, SVD5 and SVD6 focus on precise identification, making them more relevant for critical security applications. Despite SVD6's slightly lower accuracy, the targeted nature of these tasks offers a more meaningful benchmark for evaluating LLM performance in vulnerability detection.

\vspace{-0.5mm}
\begin{table}[ht]
    \centering
    \captionsetup{skip=4pt}
    \caption{LLM Performance in Identifying Vulnerabilities and Patches}
    \resizebox{\columnwidth}{!}{%
    \begin{tabular}{lcccccccc}
        \toprule
        \multirow{2}{*}{\textbf{LLMs}} & \multicolumn{2}{c}{\textbf{Precision}} & \multicolumn{2}{c}{\textbf{Recall}} & \multicolumn{2}{c}{\textbf{Accuracy}} & \multicolumn{2}{c}{\textbf{F1 Score}} \\
        \cmidrule(r){2-3} \cmidrule(l){4-5} \cmidrule(l){6-7} \cmidrule(l){8-9}
        & \textbf{SVD3,4} & \textbf{SVD5,6} & \textbf{SVD3,4} & \textbf{SVD5,6} & \textbf{SVD3,4} & \textbf{SVD5,6} & \textbf{SVD3,4} & \textbf{SVD5,6} \\
        \midrule
        \textbf{Codellama-7b}   & \cellcolor{red!0}48.95   & \cellcolor{red!20}68.08   & \cellcolor{red!0}48.53   & \cellcolor{red!12}56.95   & \cellcolor{red!0}49.26   & \cellcolor{red!25}65.15   & \cellcolor{red!0}49.02   & \cellcolor{red!8}56.10 \\
        \textbf{Codellama-34b}  & \cellcolor{red!0}49.86   & \cellcolor{red!12}60.76   & \cellcolor{red!0}49.83   & \cellcolor{red!6}54.77    & \cellcolor{red!0}51.02   & \cellcolor{red!3}52.08    & \cellcolor{red!0}51.04   & \cellcolor{red!2}51.55 \\
        \textbf{Llama3-8b}      & \cellcolor{red!0}49.65   & \cellcolor{red!65}93.16   & \cellcolor{red!0}49.35   & \cellcolor{red!20}64.78   & \cellcolor{red!0}48.50   & \cellcolor{red!50}79.15   & \cellcolor{red!0}47.56   & \cellcolor{red!25}60.15 \\
        \textbf{Llama3-70b}     & \cellcolor{red!0}47.57   & \cellcolor{red!0}28.66    & \cellcolor{red!0}48.53   & \cellcolor{red!0}35.77    & \cellcolor{red!0}46.99   & \cellcolor{red!0}28.01    & \cellcolor{red!0}48.21   & \cellcolor{red!0}35.10 \\
        \textbf{Llama3.1-8b}    & \cellcolor{red!0}50.09   & \cellcolor{red!60}88.93   & \cellcolor{red!0}50.16   & \cellcolor{red!20}64.08   & \cellcolor{red!0}49.60   & \cellcolor{red!55}80.46   & \cellcolor{red!0}49.35   & \cellcolor{red!35}61.37 \\
        \textbf{Llama3.1-70b}   & \cellcolor{red!0}50.48   & \cellcolor{red!20}68.73   & \cellcolor{red!0}50.65   & \cellcolor{red!12}58.21    & \cellcolor{red!0}49.16   & \cellcolor{red!25}66.78   & \cellcolor{red!0}48.86   & \cellcolor{red!8}56.63 \\
        \textbf{Mistral-7b}     & \cellcolor{red!0}49.38   & \cellcolor{red!3}51.47    & \cellcolor{red!0}49.35   & \cellcolor{red!0}50.40    & \cellcolor{red!0}49.59   & \cellcolor{red!0}39.41    & \cellcolor{red!0}49.67   & \cellcolor{red!0}43.92 \\
        \textbf{Mixtral-8*7b}   & \cellcolor{red!0}49.23   & \cellcolor{red!35}72.88   & \cellcolor{red!0}48.86   & \cellcolor{red!12}58.76    & \cellcolor{red!0}49.32   & \cellcolor{red!0}35.62    & \cellcolor{red!0}49.51   & \cellcolor{red!0}41.37 \\
        \textbf{Gemma2-27b}     & \cellcolor{red!8}52.33   & \cellcolor{red!0}47.56    & \cellcolor{red!3}52.12   & \cellcolor{red!0}49.83    & \cellcolor{red!0}49.59   & \cellcolor{red!15}59.61   & \cellcolor{red!0}49.51   & \cellcolor{red!8}54.14 \\
        \textbf{Gemma2-9b}      & \cellcolor{red!0}50.78   & \cellcolor{red!60}85.34   & \cellcolor{red!0}51.30   & \cellcolor{red!20}63.67   & \cellcolor{red!0}49.34   & \cellcolor{red!45}73.62   & \cellcolor{red!0}49.02   & \cellcolor{red!25}59.08 \\
        \bottomrule
    \end{tabular}%
    }
    \label{tab:is-vulnerable}
\end{table}

\subsubsection{CWE Distribution}
We analyzed the distribution of CWEs to evaluate the effectiveness of LLMs in identifying different types of vulnerabilities. The results were calculated for both SVD5 and SVD6 scenarios. The primary CWEs examined were CWE-119 (Buffer Overflow) with 50 instances, CWE-416 (Use After Free) with 37 instances, CWE-200 (Information Exposure) with 31 instances, CWE-476 (NULL Pointer Dereference) with 27 instances, and CWE-20 (Improper Input Validation) with 26 instances.
\textbf{Observations.} The Table \ref{tab:cwe-distribution} shows significant variation in model performance across different CWE types, with strengths and weaknesses emerging in handling specific vulnerabilities. Models like Llama3.1-8b and Gemma2-9b show consistently high recall, particularly with CWE-20 and CWE-200, indicating strong capabilities in identifying true vulnerabilities. In contrast, models such as Llama3-70b and Mistral-7b struggle, especially with more complex vulnerabilities like CWE-416 and CWE-476, exhibiting lower precision and recall. Codellama-7b generally performs well across most CWEs, but its effectiveness drops significantly with CWE-416. Additionally, the models tend to have lower precision, signaling a higher rate of false positives, where non-vulnerable code is mistakenly flagged as vulnerable. At the same time, the higher recall scores suggest fewer false negatives, meaning that vulnerabilities are identified effectively, though sometimes at the cost of overestimating the risk. This balance between precision and recall highlights the need for refining model accuracy to reduce false positives without sacrificing the ability to detect true vulnerabilities.

\textbf{Observations.}
The analysis of LLM performance across the five most frequent CWEs reveals key trends. A general observation is that while models displayed higher recall rates, indicating an effective detection of vulnerabilities, they often suffered from lower precision. This means that the models tended to misclassify non-vulnerable or patched code as vulnerable, leading to a significant number of false positives.

Specifically, for CWE-119 (Buffer Overflow) and CWE-416 (Use After Free), the models, such as Llama3-8b and Gemma2-9b, showed strong recall, reflecting their ability to identify these vulnerabilities. However, their lower precision and F1 scores suggest difficulties in filtering out irrelevant instances, resulting in a higher rate of false positives. Similarly, for CWE-476 (NULL Pointer Dereference), CWE-200 (Information Exposure), and CWE-20 (Improper Input Validation), the models maintained good recall but struggled with precision. This pattern across CWEs highlights the models' proficiency in detecting vulnerabilities but also emphasizes the challenge of ensuring accuracy, particularly in distinguishing true vulnerabilities from false positives. This limitation in precision undermines their practical effectiveness in scenarios where accurate vulnerability identification is critical.

\begin{tcolorbox}[before skip=5pt, after skip=5pt]
\textbf{RQ1 Summary.}
 The models exhibit significant variability in ranking vulnerabilities like CVEs and CWEs, with some showing zero Top-5 accuracy, indicating unreliable rankings. While models are better at identifying vulnerable code than confirming its safe patching, they struggle to balance recall and precision, particularly in distinguishing true vulnerabilities from benign code across different CWE types.
\end{tcolorbox}
\vspace{-0.55cm}
\begin{tcolorbox}[colback=yellow!10!white]
\textbf{RQ1 Implications.} To enhance the reliability of vulnerability detection and ranking, models need more refined training focused on systematically separating vulnerable code from its corresponding patched versions across all open-source projects. Leveraging the datasets we have developed can further enhance this process by incorporating more diverse and contextually rich examples. Improving logical reasoning, chain-of-thought processes, and precision while maintaining recall is essential. Techniques like adversarial training can help reduce false positives and improve overall decision-making in SVD tasks.
\end{tcolorbox}
\vspace{0.3mm}

\begin{table*}[ht]
\centering
\captionsetup{skip=4pt}
\caption{Model Performance Metrics by CWE Type}
\resizebox{\textwidth}{!}{%
\begin{tabular}{lcccc|cccc|cccc|cccc|cccc}
\hline
\textbf{Model} & \multicolumn{4}{c}{\textbf{CWE-119}} & \multicolumn{4}{c}{\textbf{CWE-20}} & \multicolumn{4}{c}{\textbf{CWE-200}} & \multicolumn{4}{c}{\textbf{CWE-416}} & \multicolumn{4}{c}{\textbf{CWE-476}} \\
\cline{2-21}
 & \textbf{Prec.} & \textbf{Recall} & \textbf{Acc.} & \textbf{F1} & \textbf{Prec.} & \textbf{Recall} & \textbf{Acc.} & \textbf{F1} & \textbf{Prec.} & \textbf{Recall} & \textbf{Acc.} & \textbf{F1} & \textbf{Prec.} & \textbf{Recall} & \textbf{Acc.} & \textbf{F1} & \textbf{Prec.} & \textbf{Recall} & \textbf{Acc.} & \textbf{F1} \\
\hline
Codellama-7b    & \cellcolor{red!2.94}51.47 & \cellcolor{red!40.0}70.00 & \cellcolor{red!4.0}52.00 & \cellcolor{red!18.64}59.32 
                & \cellcolor{red!5.88}52.94 & \cellcolor{red!38.46}69.23 & \cellcolor{red!7.7}53.85 & \cellcolor{red!20.0}60.00 
                & \cellcolor{red!0}47.92 & \cellcolor{red!48.38}74.19 & \cellcolor{red!0}46.77 & \cellcolor{red!16.46}58.23 
                & \cellcolor{red!0}42.86 & \cellcolor{red!0}48.65 & \cellcolor{red!0}41.89 & \cellcolor{red!0}45.57 
                & \cellcolor{red!0}50.00 & \cellcolor{red!25.92}62.96 & \cellcolor{red!0}50.00 & \cellcolor{red!11.48}55.74 \\
Codellama-34b   & \cellcolor{red!6.12}53.06 & \cellcolor{red!10.64}55.32 & \cellcolor{red!6.38}53.19 & \cellcolor{red!8.34}54.17 
                & \cellcolor{red!4.34}52.17 & \cellcolor{red!4.34}52.17 & \cellcolor{red!4.34}52.17 & \cellcolor{red!4.34}52.17 
                & \cellcolor{red!12.0}56.00 & \cellcolor{red!0}45.16 & \cellcolor{red!9.68}54.84 & \cellcolor{red!0}50.00 
                & \cellcolor{red!0}47.50 & \cellcolor{red!5.56}52.78 & \cellcolor{red!0}47.22 & \cellcolor{red!0}50.00 
                & \cellcolor{red!0}46.15 & \cellcolor{red!0}46.15 & \cellcolor{red!0}46.15 & \cellcolor{red!0}46.15 \\
Llama3-8b       & \cellcolor{red!0}46.34 & \cellcolor{red!50}76.00 & \cellcolor{red!0}44.00 & \cellcolor{red!15.16}57.58 
                & \cellcolor{red!0}45.24 & \cellcolor{red!46.16}73.08 & \cellcolor{red!0}42.31 & \cellcolor{red!11.76}55.88 
                & \cellcolor{red!1.96}50.98 & \cellcolor{red!50}83.87 & \cellcolor{red!3.22}51.61 & \cellcolor{red!26.82}63.41 
                & \cellcolor{red!0}49.12 & \cellcolor{red!50}75.68 & \cellcolor{red!0}48.65 & \cellcolor{red!19.14}59.57 
                & \cellcolor{red!0}50.00 & \cellcolor{red!48.14}74.07 & \cellcolor{red!0}50.00 & \cellcolor{red!19.4}59.7 \\
Llama3-70b      & \cellcolor{red!0}48.72 & \cellcolor{red!0}38.00 & \cellcolor{red!0}49.00 & \cellcolor{red!0}42.70 
                & \cellcolor{red!5.26}52.63 & \cellcolor{red!0}38.46 & \cellcolor{red!3.84}51.92 & \cellcolor{red!0}44.44 
                & \cellcolor{red!0}38.46 & \cellcolor{red!0}16.13 & \cellcolor{red!0}45.16 & \cellcolor{red!0}22.73 
                & \cellcolor{red!0}41.18 & \cellcolor{red!0}18.92 & \cellcolor{red!0}45.95 & \cellcolor{red!0}25.93 
                & \cellcolor{red!0}40.00 & \cellcolor{red!0}14.81 & \cellcolor{red!0}46.30 & \cellcolor{red!0}21.62 \\
Llama3.1-8b     & \cellcolor{red!0}47.44 & \cellcolor{red!48}74.00 & \cellcolor{red!0}46.00 & \cellcolor{red!15.62}57.81 
                & \cellcolor{red!6.38}53.19 & \cellcolor{red!50}96.15 & \cellcolor{red!11.54}55.77 & \cellcolor{red!36.98}68.49 
                & \cellcolor{red!8.70}54.35 & \cellcolor{red!50}80.65 & \cellcolor{red!12.9}56.45 & \cellcolor{red!29.88}64.94 
                & \cellcolor{red!0}50.00 & \cellcolor{red!50}75.68 & \cellcolor{red!0}50.00 & \cellcolor{red!20.44}60.22 
                & \cellcolor{red!0}48.89 & \cellcolor{red!50}81.48 & \cellcolor{red!0}48.15 & \cellcolor{red!22.22}61.11 \\
Llama3.1-70b    & \cellcolor{red!0}48.53 & \cellcolor{red!32.0}66.00 & \cellcolor{red!0}48.00 & \cellcolor{red!11.86}55.93 
                & \cellcolor{red!5.0}52.50 & \cellcolor{red!50}80.77 & \cellcolor{red!7.7}53.85 & \cellcolor{red!27.28}63.64 
                & \cellcolor{red!0}37.84 & \cellcolor{red!0}45.16 & \cellcolor{red!0}35.48 & \cellcolor{red!0}41.18 
                & \cellcolor{red!2.04}51.02 & \cellcolor{red!35.14}67.57 & \cellcolor{red!2.7}51.35 & \cellcolor{red!16.28}58.14 
                & \cellcolor{red!6.24}53.12 & \cellcolor{red!25.92}62.96 & \cellcolor{red!7.4}53.70 & \cellcolor{red!15.26}57.63 \\
Mistral-7b      & \cellcolor{red!0}48.78 & \cellcolor{red!0}40.00 & \cellcolor{red!0}49.00 & \cellcolor{red!0}43.96 
                & \cellcolor{red!0}40.91 & \cellcolor{red!0}34.62 & \cellcolor{red!0}42.31 & \cellcolor{red!0}37.50 
                & \cellcolor{red!0}40.00 & \cellcolor{red!0}32.26 & \cellcolor{red!0}41.94 & \cellcolor{red!0}35.71 
                & \cellcolor{red!0}46.67 & \cellcolor{red!0}37.84 & \cellcolor{red!0}47.30 & \cellcolor{red!0}41.79 
                & \cellcolor{red!12.5}56.25 & \cellcolor{red!0}33.33 & \cellcolor{red!7.4}53.70 & \cellcolor{red!0}41.86 \\
Mixtral-8*7b    & \cellcolor{red!0}50.00 & \cellcolor{red!0}36.73 & \cellcolor{red!0}50.00 & \cellcolor{red!0}42.35 
                & \cellcolor{red!8.34}54.17 & \cellcolor{red!0}50.00 & \cellcolor{red!7.7}53.85 & \cellcolor{red!4.0}52.00 
                & \cellcolor{red!4.34}52.17 & \cellcolor{red!0}38.71 & \cellcolor{red!3.22}51.61 & \cellcolor{red!0}44.44 
                & \cellcolor{red!11.12}55.56 & \cellcolor{red!0}40.54 & \cellcolor{red!8.1}54.05 & \cellcolor{red!0}46.88 
                & \cellcolor{red!10.0}55.00 & \cellcolor{red!0}40.74 & \cellcolor{red!7.4}53.70 & \cellcolor{red!0}46.81 \\
Gemma2-9b       & \cellcolor{red!5.12}52.56 & \cellcolor{red!50}82.00 & \cellcolor{red!8.0}54.00 & \cellcolor{red!28.12}64.06 
                & \cellcolor{red!6.24}53.12 & \cellcolor{red!30.76}65.38 & \cellcolor{red!7.7}53.85 & \cellcolor{red!17.24}58.62 
                & \cellcolor{red!0}48.21 & \cellcolor{red!50}87.10 & \cellcolor{red!0}46.77 & \cellcolor{red!24.14}62.07 
                & \cellcolor{red!3.7}51.85 & \cellcolor{red!50}75.68 & \cellcolor{red!5.4}52.70 & \cellcolor{red!23.08}61.54 
                & \cellcolor{red!0}45.71 & \cellcolor{red!18.52}59.26 & \cellcolor{red!0}44.44 & \cellcolor{red!3.22}51.61 \\
Gemma2-27b      & \cellcolor{red!0}49.30 & \cellcolor{red!40.0}70.00 & \cellcolor{red!0}49.00 & \cellcolor{red!15.70}57.85 
                & \cellcolor{red!0}50.00 & \cellcolor{red!7.7}53.85 & \cellcolor{red!0}50.00 & \cellcolor{red!3.7}51.85 
                & \cellcolor{red!0}42.42 & \cellcolor{red!0}45.16 & \cellcolor{red!0}41.94 & \cellcolor{red!0}43.75 
                & \cellcolor{red!0}47.37 & \cellcolor{red!0}48.65 & \cellcolor{red!0}47.30 & \cellcolor{red!0}48.00 
                & \cellcolor{red!11.12}55.56 & \cellcolor{red!11.12}55.56 & \cellcolor{red!11.12}55.56 & \cellcolor{red!11.12}55.56 \\
\hline
\end{tabular}%
}
\label{tab:cwe-distribution}
\end{table*}

% ###########################################RQ2############################################
\subsection{RQ2: Effectiveness of LLMs in SVP}

To understand the capabilities of LLMs on SVP, we calculated ROUGE scores to assess the similarity between generated and reference patches, CodeBLEU scores to evaluate the quality of the generated code, and cyclomatic complexity to measure the complexity of the code. We considered both zero-shot and few-shot learning scenarios to observe how commit description affects the models' performance. 

\begin{figure}[htbp]
\centerline{\includegraphics[width=0.5\textwidth]{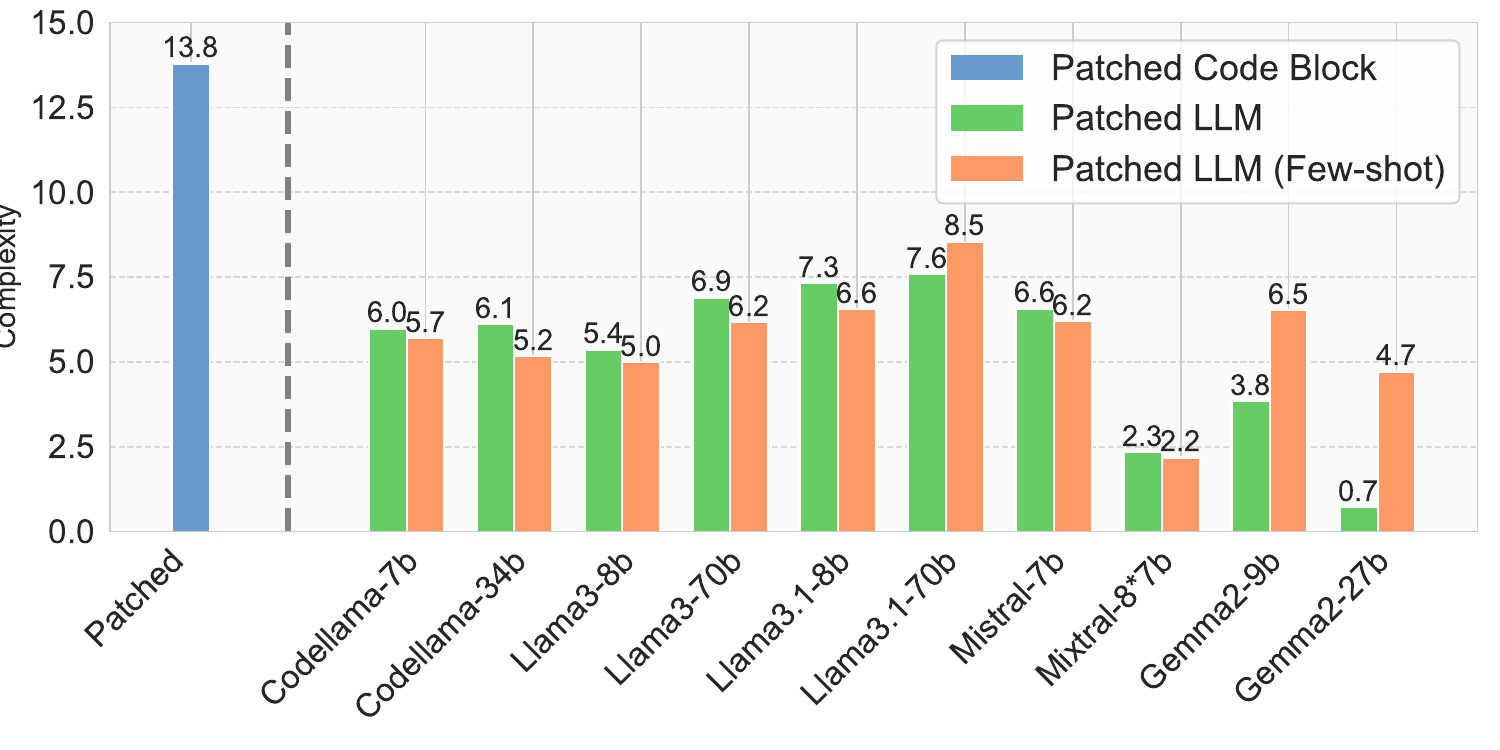}}
\caption{ Complexity Scores}
\label{figure:complexities}
\end{figure}

\textbf{Observations.} 
The cyclomatic complexity analysis, illustrated in Figure \ref{figure:complexities}, shows that the LLMs consistently generated code with lower complexity compared to the original vulnerable code. This trend was observed across almost all models, both in zero-shot (Z) and few-shot (F) settings. Notably, while reducing complexity can lead to more maintainable code, it also suggests that the LLMs might be simplifying the code too much, potentially overlooking important details required for a robust and secure patch. The simplified patches often lacked the depth needed to address the vulnerabilities comprehensively, which is a critical concern in SVP tasks. Additionally, the models exhibited a tendency to simplify the structure of the code without fully addressing the underlying logic, as indicated by placeholders like "..." or "the rest of the code remains unchanged," in their outputs.

\begin{figure*}[htbp]
\centerline{\includegraphics[width=1\textwidth]{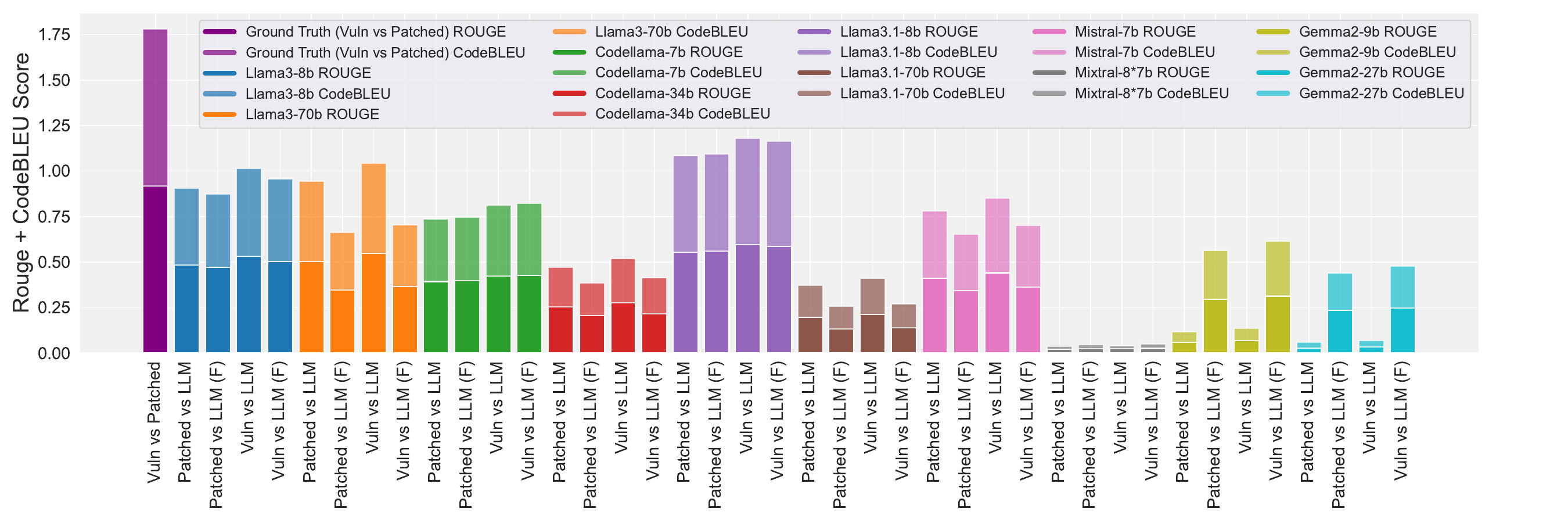}}
\caption{Similarity Scores (ROUGE Scores, CodeBLEU Scores)}
\label{figure:similarities}
\end{figure*}

Figure \ref{figure:similarities} offers a comparative view of the ROUGE and CodeBLEU scores across different models, highlighting the varying degrees of similarity and quality in the generated patches. The scores reveal that although some LLMs perform relatively well in producing code that is syntactically accurate, they frequently miss critical contextual elements, which are essential for creating functionally sound and secure patches. This discrepancy underscores the importance of context-aware generation, where the model not only reproduces the structure of the code but also understands and integrates the specific requirements of the vulnerability being addressed.

\begin{tcolorbox}[before skip=5pt, after skip=5pt]
\textbf{RQ2 Summary.}
LLMs consistently generated code with lower cyclomatic complexity, the simplified patches often lacked the depth required to fully address vulnerabilities. ROUGE and CodeBLEU scores further revealed that the models frequently missed critical contextual elements.
\end{tcolorbox}
\vspace{-0.65cm}
\begin{tcolorbox}[colback=yellow!10!white]
\textbf{RQ2 Implications.} To enhance the quality of generated patches, LLMs should adhere closely to the context provided by users and generate complete patches rather than partial fixes. Improving the models' ability to fully understand the context and encouraging them to produce patches that closely match the length and complexity of the original vulnerable code is essential.
\end{tcolorbox}
% ##########################################RQ3#############################################
\subsection{RQ3: Impact of Code Block Abstraction Level and Length}
In this section, we examine how the number of files and functions within a vulnerable code block affects the accuracy of LLMs in SVD tasks. Additionally, we analyze how the number of lines in these code blocks influences the ability to accurately identify vulnerabilities. This analysis helps us understand how complexity and size of the code impact the performance of LLMs in detecting vulnerabilities.

\subsubsection{Abstraction Levels}
The abstraction level of a code block, defined by the number of files and functions it spans, plays a critical role in determining the effectiveness of LLMs in SVD tasks. We assess the abstraction level by examining both the number of files affected in a commit patch and the number of functions within vulnerable code blocks, which are extracted using regular expressions. This abstraction level—ranging from a single file with a single function to multiple files with multiple functions—can significantly influence the model's effectiveness. In this analysis, we explore how LLMs respond to these varying levels of code block abstraction, analyzing their accuracy, precision, recall, and F1 scores across different scenarios.

\textbf{Observations.} As shown in Table \ref{tab:abstraction}, the effectiveness of LLMs across different abstraction levels—ranging from Level 1 to Level 3—exhibits noticeable variability and non-deterministic behavior. Using prompts SVD5 and SVD6 from Table\ref{table:prompt_templates}, we observed that Llama3.1-8b consistently outperformed other models at all levels of abstraction. At Level 1, involving a single file and function, it achieved a recall of 75.18\% and an F1 score of 59.72, indicating strong performance in detecting vulnerabilities. Gemma2-9b exhibited similar performance in this category. As the complexity increased to Level 2, which involves one file with multiple functions, Llama3.1-8b’s recall improved to 86.79\%, with an F1 score of 63.01, suggesting that additional functions provide useful context for vulnerability detection. However, at the most complex Level 3, where multiple files and functions are involved, there was a slight decrease in performance, reflecting the challenges LLMs face when dealing with more extensive and intricate datasets. Despite this, Llama3.1-8b still maintained strong recall and F1 scores, underscoring its robustness across varying levels of abstraction.

\begin{table*}[ht]
\centering
\captionsetup{skip=4pt}
\caption{Model Performance Metrics by Abstraction Level}
% \scriptsize
\resizebox{\textwidth}{!}{%
\begin{tabular}{lcccc|cccc|cccc}
\hline
\textbf{Model} & \multicolumn{4}{c}{\textbf{Level 1}} & \multicolumn{4}{c}{\textbf{Level 2}} & \multicolumn{4}{c}{\textbf{Level 3}} \\
\cline{2-5} \cline{6-9} \cline{10-13}
& \textbf{Prec.} & \textbf{Recall} & \textbf{Acc.} & \textbf{F1} & \textbf{Prec.} & \textbf{Recall} & \textbf{Acc.} & \textbf{F1} & \textbf{Prec.} & \textbf{Recall} & \textbf{Acc.} & \textbf{F1} \\
\hline
Codellama-34b   & \cellcolor{red!0.00}48.44 & \cellcolor{red!0.00}46.62 & \cellcolor{red!0.00}47.71 & \cellcolor{red!0.00}47.51 
                & \cellcolor{red!14.28}57.14 & \cellcolor{red!25.50}62.75 & \cellcolor{red!14.00}57.00 & \cellcolor{red!19.62}59.81 
                & \cellcolor{red!5.00}52.50 & \cellcolor{red!5.88}52.94 & \cellcolor{red!3.82}51.91 & \cellcolor{red!5.44}52.72 \\
Codellama-7b    & \cellcolor{red!0.00}48.39 & \cellcolor{red!48.94}74.47 & \cellcolor{red!0.00}47.52 & \cellcolor{red!17.32}58.66 
                & \cellcolor{red!0.00}46.43 & \cellcolor{red!0.00}49.06 & \cellcolor{red!0.00}46.23 & \cellcolor{red!0.00}47.71 
                & \cellcolor{red!3.76}51.88 & \cellcolor{red!22.12}61.06 & \cellcolor{red!4.42}52.21 & \cellcolor{red!12.20}56.10 \\
Gemma2-27b      & \cellcolor{red!0.00}48.25 & \cellcolor{red!0.00}48.94 & \cellcolor{red!0.00}48.23 & \cellcolor{red!0.00}48.59 
                & \cellcolor{red!0.00}49.18 & \cellcolor{red!13.20}56.60 & \cellcolor{red!0.00}49.06 & \cellcolor{red!5.26}52.63 
                & \cellcolor{red!1.82}50.91 & \cellcolor{red!48.68}74.34 & \cellcolor{red!2.66}51.33 & \cellcolor{red!20.86}60.43 \\
Gemma2-9b       & \cellcolor{red!0.48}50.24 & \cellcolor{red!48.94}74.47 & \cellcolor{red!0.70}50.35 & \cellcolor{red!20.00}60.00 
                & \cellcolor{red!0.00}49.30 & \cellcolor{red!32.08}66.04 & \cellcolor{red!0.00}49.06 & \cellcolor{red!12.90}56.45 
                & \cellcolor{red!0.00}48.31 & \cellcolor{red!50.00}76.11 & \cellcolor{red!0.00}47.35 & \cellcolor{red!18.22}59.11 \\
Llama3-70b      & \cellcolor{red!0.00}46.67 & \cellcolor{red!0.00}14.89 & \cellcolor{red!0.00}48.94 & \cellcolor{red!0.00}22.58 
                & \cellcolor{red!0.00}39.47 & \cellcolor{red!0.00}28.30 & \cellcolor{red!0.00}42.45 & \cellcolor{red!0.00}32.97 
                & \cellcolor{red!0.00}50.00 & \cellcolor{red!0.00}44.25 & \cellcolor{red!0.00}50.00 & \cellcolor{red!0.00}46.95 \\
Llama3-8b       & \cellcolor{red!0.00}47.32 & \cellcolor{red!50.00}75.18 & \cellcolor{red!0.00}45.74 & \cellcolor{red!16.16}58.08 
                & \cellcolor{red!5.88}52.94 & \cellcolor{red!50.00}84.91 & \cellcolor{red!9.44}54.72 & \cellcolor{red!30.44}65.22 
                & \cellcolor{red!0.00}47.92 & \cellcolor{red!50.00}81.42 & \cellcolor{red!0.00}46.46 & \cellcolor{red!20.66}60.33 \\
Llama3.1-70b    & \cellcolor{red!0.00}50.00 & \cellcolor{red!23.40}61.70 & \cellcolor{red!0.00}50.00 & \cellcolor{red!10.48}55.24 
                & \cellcolor{red!0.00}48.68 & \cellcolor{red!39.62}69.81 & \cellcolor{red!0.00}48.11 & \cellcolor{red!14.72}57.36 
                & \cellcolor{red!0.00}48.50 & \cellcolor{red!43.36}71.68 & \cellcolor{red!0.00}47.79 & \cellcolor{red!15.72}57.86 \\
Llama3.1-8b     & \cellcolor{red!0.00}49.53 & \cellcolor{red!50.00}75.18 & \cellcolor{red!0.00}49.29 & \cellcolor{red!19.44}59.72 
                & \cellcolor{red!0.00}49.46 & \cellcolor{red!50.00}86.79 & \cellcolor{red!0.00}49.06 & \cellcolor{red!26.02}63.01 
                & \cellcolor{red!0.00}49.74 & \cellcolor{red!50.00}84.07 & \cellcolor{red!0.00}49.56 & \cellcolor{red!25.00}62.50 \\
Mistral-7b      & \cellcolor{red!0.00}44.66 & \cellcolor{red!0.00}32.62 & \cellcolor{red!0.00}46.10 & \cellcolor{red!0.00}37.70 
                & \cellcolor{red!11.76}55.88 & \cellcolor{red!0.00}35.85 & \cellcolor{red!7.54}53.77 & \cellcolor{red!0.00}43.68 
                & \cellcolor{red!4.68}52.34 & \cellcolor{red!0.00}49.56 & \cellcolor{red!4.42}52.21 & \cellcolor{red!1.82}50.91 \\
Mixtral-8*7b    & \cellcolor{red!0.00}48.24 & \cellcolor{red!0.00}29.08 & \cellcolor{red!0.00}48.94 & \cellcolor{red!0.00}36.28 
                & \cellcolor{red!2.86}51.43 & \cellcolor{red!0.00}33.96 & \cellcolor{red!0.96}50.48 & \cellcolor{red!0.00}40.91 
                & \cellcolor{red!0.00}50.00 & \cellcolor{red!0.00}45.13 & \cellcolor{red!0.00}50.00 & \cellcolor{red!0.00}47.44 \\
\hline
\end{tabular}%
}
\begin{flushleft}
\centering
\textbf{Level 1:} 1 file, 1 function, \textbf{Level 2:} 1 file, multiple functions, \textbf{Level 3:} multiple files, multiple functions
\end{flushleft}
\label{tab:abstraction}
\end{table*}

\subsubsection{Impact of Code Length}
The length of a code block, measured by the number of lines it contains, directly impacts the performance of LLMs in detecting vulnerabilities. Longer code blocks can introduce challenges related to context retention and increased complexity, while shorter blocks may lack sufficient information for accurate analysis. In this section, we explore how different code lengths affect LLM performance metrics providing a detailed understanding of the models' strengths and weaknesses when dealing with varying amounts of code.

\begin{figure*}[htbp]
\centerline{\includegraphics[width=1\textwidth]{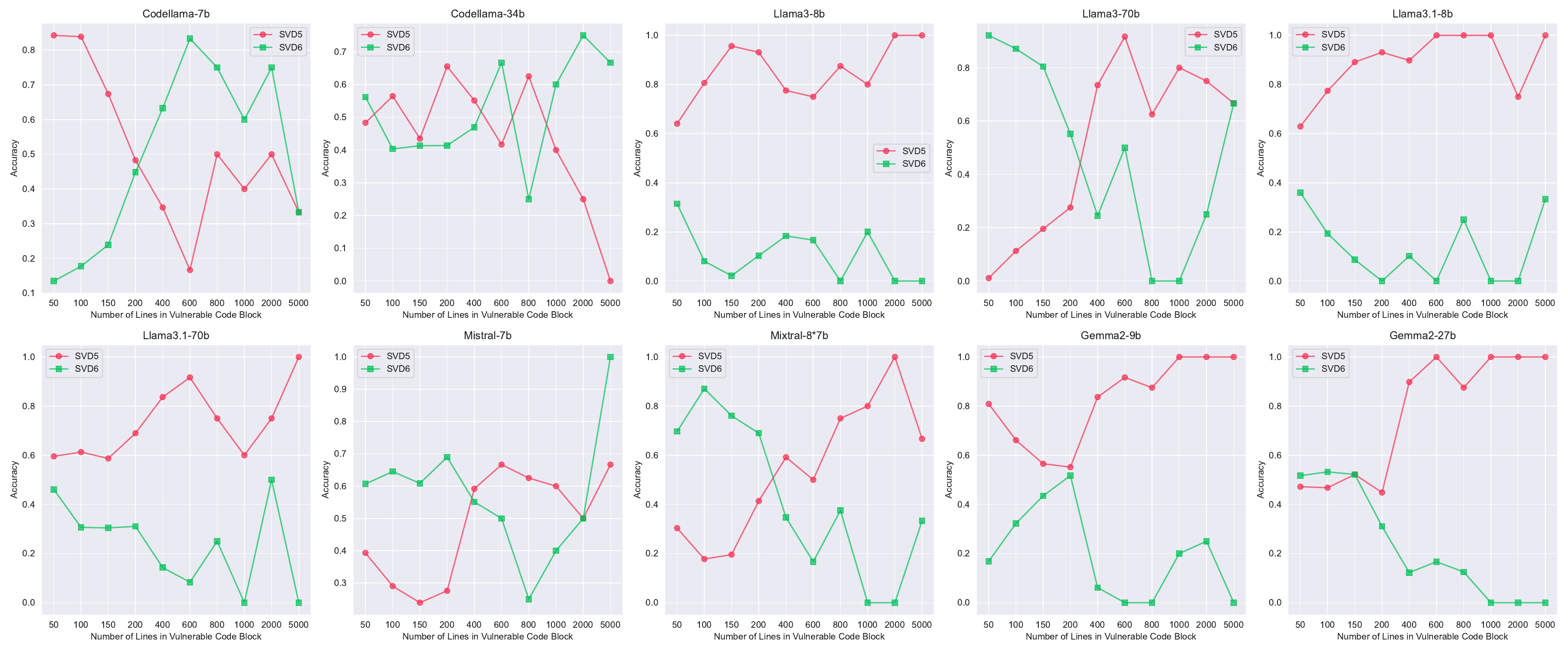}}
\caption{Accuracy of LLMs Based on the Number of Lines in Vulnerable Code Blocks}
\label{figure:acc-code-length}
\end{figure*}

\textbf{Observations.} SVD5, which checks whether a vulnerable code block corresponds to its associated CVE and CWE, typically shows higher accuracy across models, as it works with more distinct patterns. On the other hand, SVD6, which performs the same verification on patched code, encounters difficulties due to the modifications introduced in the patches, leading to lower accuracy. This suggests that LLMs are more effective at identifying vulnerabilities in original code compared to patched versions. The performance variations in SVD6 across different models highlight the need for enhanced methods or additional context when analyzing patched code.
Interestingly, Figure \ref{figure:acc-code-length} does not reveal a clear correlation between code length and detection performance. Some models show improvement as the number of lines in the vulnerable code block increases, while others exhibit declining or inconsistent performance. This inconsistency suggests that the length of the code alone is not a reliable indicator of the LLM's effectiveness in detecting vulnerabilities or verifying patches.

\begin{figure*}[t!]
\centerline{\includegraphics[width=1\textwidth]{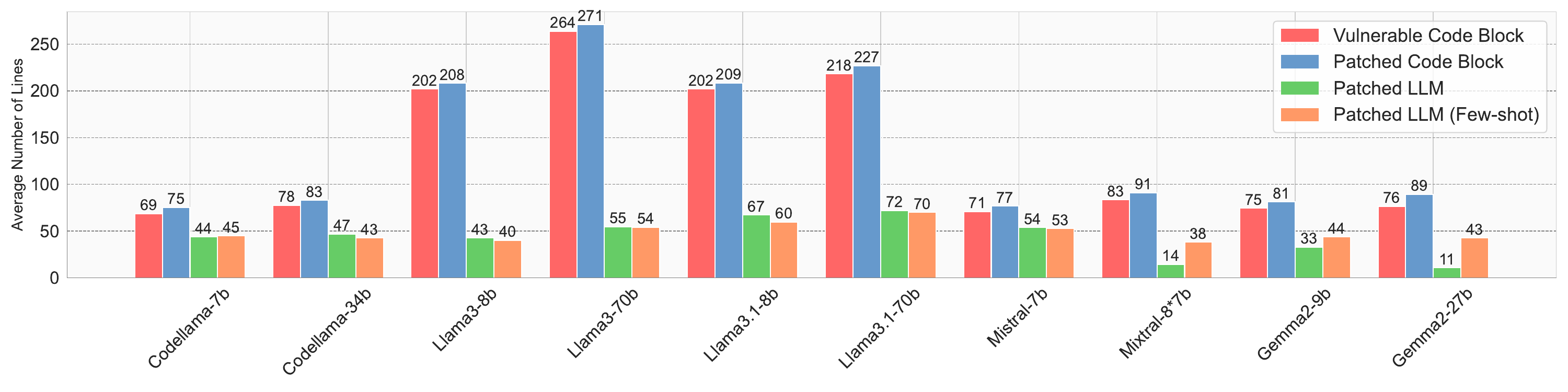}}
\caption{Average Lines of Code in Extracted Vulnerable and Patched Code Blocks vs. Code Blocks Generated by LLMs}
\label{figure:average-lines-of-code}
\end{figure*}

Figure \ref{figure:average-lines-of-code} presents the average length of vulnerable, patched, LLM-generated patched, and few-shot LLM-generated patched code blocks. The preprocessing and filtering steps to exclude blank LLM code blocks resulted in different subsets of data for each model, leading to variations in the average line counts. Notably, the average length of LLM-generated patches is consistently shorter than the original patches, which indicates that LLMs tend to produce more concise code modifications. This reduction in code length might suggest a focus on essential changes, but it could also imply that LLMs struggle to fully replicate the complexity of the original patches.

\begin{tcolorbox}[before skip=5pt, after skip=5pt]
\textbf{RQ3 Summary.}
The analysis shows that LLMs perform inconsistently across code abstraction levels and lengths, excelling in SVD3 and SVD5 tasks. They improve with more files and functions, leveraging larger contexts. However, LLM-generated patches tend to be shorter, highlighting difficulties in replicating complex code modifications.
\end{tcolorbox}
\vspace{-0.6cm}
\begin{tcolorbox}[colback=yellow!10!white]
\textbf{RQ3 Implications.}
Given the large context windows of most LLMs, refining training data to include more complex and multi-function code examples could help the models better utilize their contextual capabilities, leading to improved accuracy in handling intricate code structures.
\end{tcolorbox}

\section{Discussion and Research Directions}
\label{section:discussion}
VulnLLMEval represents a significant advancement in the evaluation of LLMs for automated SVD and SVP. By providing a comprehensive framework for assessing LLM capabilities across a range of tasks, our work opens new avenues for research in this critical area of cybersecurity. This section discusses the challenges and limitations encountered during our study, providing insights into areas where further improvements are needed, and outlines future research directions.
\subsection{Challenges and Limitations}
\textit{Data Bias and Model Variability.}
To mitigate bias introduced by models that may have been inadvertently trained on similar datasets, we created a custom dataset ensuring data uniqueness. This dataset, which can be extended with other programming languages and open-source projects, was essential in reducing potential skewing of results. However, LLMs exhibited inconsistencies in CVE and CWE rankings due to their non-deterministic nature. Despite efforts to stabilize outputs using consistent temperature settings, variability persisted, particularly in balancing recall and precision for different vulnerability types.\\
\textit{CWE Complexity and Evaluation Metrics.}
Certain CWE types, like CWE-416, presented significant challenges, especially in the data-flow match component of the CodeBLEU metric. This was particularly evident when code blocks lacked functions or consisted mainly of non-functional elements, leading to incomplete evaluations. These difficulties underscore the need for functionally rich code samples in assessments, as their absence may prevent models from fully capturing the nuances of specific vulnerabilities.\\
\textit{Limitations in Real-World Application.}
Despite our efforts, the LLMs often produced oversimplified patches that failed to address vulnerabilities comprehensively. In addition, in SVP tasks, some LLMs tended to produce empty responses, which posed a significant challenge in generating usable solutions. Challenges in evaluating complex CWE types, such as CWE-416, highlighted the difficulty of achieving accurate code analysis using existing metrics. While our study primarily focused on benchmarking and evaluation, it opens the door to further research in areas like prompt engineering and advanced techniques such as functionality testing with Pass@K \cite{xu2022systematic, chen2021evaluating}. These techniques, although not the focus of our current work, hold promise for significantly improving LLM performance in software vulnerability detection and patching.

\subsection{Research Directions}
Future research in automated SVD and SVP with LLMs should prioritize developing models that can accurately differentiate between vulnerable and patched code. Training LLMs on general code alone is insufficient; instead, targeted fine-tuning, as illustrated by the data collection and labeling techniques in this study, is crucial for improving the precision of these models in identifying and mitigating security vulnerabilities \cite{xia2023automated}. 

While it is important to continuously adapt LLMs to new vulnerabilities, we must also address the challenge of memorization. Over time, as LLMs are exposed to new data, they tend to forget previously learned information—a phenomenon known as catastrophic forgetting \cite{goodfellow2013empirical}. This issue is particularly problematic in SVD and SVP tasks, where retaining knowledge of past vulnerabilities and patches is critical for effective detection and remediation of security issues. Future research should explore strategies to mitigate the effects of memorization in LLMs, such as meta-learning and Elastic Weight Consolidation (EWC), to ensure that these models remain reliable and effective as they evolve \cite{vilalta2002perspective, kirkpatrick2017overcoming, hospedales2021meta}.

Although LLMs have shown promise in tasks like code completion and syntax highlighting, their current limitations in dynamic decision-making and multi-step reasoning restrict their effectiveness in automated SVD and SVP \cite{zibaeirad2024comprehensive, yao2022react, wang2024executable, shentu2024llms}. To overcome these challenges, future efforts should enhance LLMs' abilities to grasp the context of vulnerabilities, detect intricate patterns, and apply accurate code patches. This will require models that extend beyond simple action-based responses, integrating reasoning processes that allow for informed and effective vulnerability management.

Finally, combining LLMs with other AI approaches, such as reinforcement learning and hybrid neural networks, could significantly improve their capacity to automate key aspects of software security. The integration of agents that unify reasoning with action, as proposed in recent studies, offers a promising path forward to address the current limitations of LLMs in this field \cite{park2023generative}.

\section{Threads to Validity}
\label{section:threadstovalidity}

\textbf{Internal Validity.}
The non-deterministic nature of LLMs posed a significant challenge, as their outputs can vary between runs. To mitigate this, we set the temperature to 0, ensuring more stable and consistent results across both SVD and SVP tasks. This stability allowed us to reduce output variability, though minor variations might still occur. Additionally, we avoided the potential for human error by eliminating manual labeling, relying on real-world vulnerabilities from our comprehensive dataset. Our decision to use real-world data reflects the inherent complexity and interconnectedness of vulnerabilities, providing a more reliable evaluation. However, limitations remain due to the complexity of differentiating between vulnerable and patched code, as models often struggled with this distinction.\\
\textbf{External Validity.}
We utilized various LLMs with different variants to ensure the portability of our evaluation. While our evaluation focused on the Linux kernel, which limits generalizability to other systems or languages, the insights gained are applicable to broader SVD and SVP challenges. The models we tested, varying in parameters and purposes (e.g., code generation or instruction following), demonstrated that the underlying characteristics of the codebase (C language) and the specific structure of the Linux kernel might not directly apply to other projects. The generalizability of the results to other programming environments remains an open question, highlighting a need for future research across different languages and systems.\\
\textbf{Construct Validity.}
Potential bugs in the evaluation framework or model-generated code could introduce bias into our results. To minimize this risk, we conducted peer reviews of the evaluation process, ensuring the accuracy of our framework. Additionally, we continuously refined our filtering mechanisms to reduce manual oversight and improve consistency in how model outputs were labeled and evaluated.\\
\textbf{Statistical Conclusion Validity.}
Our use of consistent metrics (e.g., ROUGE, CodeBLEU, cyclomatic complexity) across all models ensured comparability, and the inclusion of a large, diverse dataset captured the variability of real-world vulnerabilities. Sensitivity analyses and robustness checks were performed to ensure the stability and reliability of the findings, further supporting the validity of our conclusions.\\

\section{Conclusion}
\label{section:conclusion}
This paper introduced VulnLLMEval, a comprehensive benchmarking framework and dataset designed to evaluate the effectiveness of LLMs in SVD and SVP. Our approach automates dataset collection, focusing on real-world vulnerabilities from the Linux kernel, to ensure consistent and realistic evaluations. The results reveal that while LLMs perform well in detecting vulnerabilities, they often struggle to differentiate accurately between vulnerable and patched code, sometimes oversimplifying patches. The study also shows that LLM performance improves with increased code context but declines with added complexity, offering valuable insights for enhancing LLMs in software security tasks. Finally, we propose research directions to address these challenges, including mitigating catastrophic forgetting in LLMs through meta-learning, and integrating reinforcement learning to improve dynamic decision-making in software vulnerability detection and patching.

\bibliographystyle{ACM-Reference-Format}
\bibliography{ref}
% \section{Appendix}
% figures for lines < 150 (for svd and svp)

\end{document}